\documentclass[11pt,a4paper]{article}
\pdfoutput=1
\usepackage{textcomp}
\usepackage{newunicodechar}
\newunicodechar{ỳ}{\`y}

\usepackage{amsmath}
\usepackage{mathtools}          
\usepackage{amsfonts}
\usepackage{amssymb}
\usepackage{latexsym}
\usepackage[nointegrals]{wasysym}   
\usepackage{mathrsfs}            
\usepackage{slashed}
\usepackage{nicefrac}
\usepackage{accents}
\usepackage{braket}
\usepackage{amsthm}

\usepackage[dvipsnames,svgnames,table]{xcolor}

\definecolor{dark-gray}{gray}{0.20}
\definecolor{medium-gray}{gray}{0.30}     
\definecolor{light-gray}{gray}{0.80}
\definecolor{dark-red}{rgb}{0.7,0,0}
\definecolor{dark-green}{rgb}{0.1,0.4,0}
\definecolor{dark-blue}{rgb}{0.3,0.3,0.7}
\definecolor{light-blue}{rgb}{0.8,0.8,1}
\definecolor{swamp}{RGB}{240,199,197}
\definecolor{landscape}{RGB}{180,250,199}
\definecolor{undecided}{RGB}{252,252,197}

\usepackage{array}
\usepackage{multirow}
\usepackage{tabularx}
\usepackage{colortbl}
\usepackage{longtable}
\usepackage{arydshln}

\usepackage{graphicx}
\usepackage{graphbox}           
\usepackage{epsfig}           
\usepackage{float}
\usepackage{wrapfig}
\usepackage{caption}
\usepackage{subcaption}

\usepackage{tikz}
\usepackage{tikz-cd}
\usepackage{tikz-3dplot}
\usetikzlibrary{shapes.geometric}
\usetikzlibrary{arrows}
\usetikzlibrary{3d}

\usepackage{setspace}
\usepackage{titlesec}
\usepackage{fancyhdr}

\usepackage[shortlabels]{enumitem} 
\usepackage[most]{tcolorbox}
\usepackage{soul}
\usepackage{rotate}
\usepackage{pifont}
\usepackage{comment}

\usepackage{jheppub}     
\usepackage{hyperref}   

\hypersetup{
    colorlinks   = true,
    linkcolor    = dark-blue,
    citecolor    = dark-red,
    urlcolor     = dark-green,
    filecolor    = Blue,
    linktoc      = page,
    pdfstartview = FitV,
    bookmarksopen= true
}

\definecolor{pastelyellow}{RGB}{255, 245, 170}

\numberwithin{equation}{section}

\theoremstyle{plain}

\theoremstyle{remark}

\newcommand{\be}{\begin{equation}}
\newcommand{\ee}{\end{equation}}
\newcommand{\beq}{\begin{equation}}
\newcommand{\eeq}{\end{equation}}
\newcommand{\bal}{\begin{aligned}}
\newcommand{\eal}{\end{aligned}}
\newcommand{\bea}{\begin{eqnarray}}   
\newcommand{\eea}{\end{eqnarray}}

\newcommand{\s}{\text{s}}
\renewcommand{\t}{\text{t}}
\renewcommand{\b}{\text{b}}
\newcommand{\n}{$\mathcal{N}=1 \ $}

\newcommand{\R}{\mathbb{R}}

\newcommand{\Mpf}{M_{\textrm{P,4}}}

\newcommand{\abs}[1]{\lvert #1 \rvert}

\newcommand{\simleq}{\; \raise0.3ex\hbox{$<$\kern-0.75em
                        \raise-1.1ex\hbox{$\sim$}}\; }
\newcommand{\simgeq}{\; \raise0.3ex\hbox{$>$\kern-0.75em
                        \raise-1.1ex\hbox{$\sim$}}\; }

\newcommand{\figref}[1]{figure~\ref{#1}}

\title{\centering Light EFT Strings and \\Asymptotic Accelerated Expansion}

\author{Veronica Collazuol$^{1,2}$, Alessandra Grieco$^{1,2}$}\affiliation{$^{1}$Instituto de F\'{i}sica Te\'{o}rica IFT-UAM/CSIC,
C/ Nicol\'{a}s Cabrera 13-15, Campus de Cantoblanco, 28049 Madrid, Spain}
\affiliation{$^2$Departamento de F\'{i}sica Te\'{o}rica, Universidad Aut\'{o}noma de Madrid, Cantoblanco, 28049 Madrid, Spain}
\author{and Irene Valenzuela$^{1,2,3}$}\affiliation{$^3$CERN, Theoretical Physics Department, 1211 Meyrin, Switzerland}

\abstract{We study the interplay among towers of particles, EFT strings, and EFT membranes near perturbative infinite-distance limits of four-dimensional $\mathcal{N}=1$ theories. We show that the scaling vectors of EFT membranes lie in the lattice generated by those of the EFT strings, as previously found for particle towers, and determine their specific relative scaling in F-theory and Type IIA Calabi–Yau compactifications. We then leverage these objects to assess the viability of asymptotic accelerated expansion in flux compactifications. While our bottom-up analysis does not yield a general obstruction, we rule out two candidate accelerating trajectories in F-theory because an EFT string becomes lighter than the Hubble scale, invalidating the four-dimensional EFT. This closes the loophole found in arXiv:2209.11821. We also study how recent results on quantum obstructions to $\mathcal{N}=1$ infinite-distance limits may provide a concrete resolution by changing the asymptotic form of the K\"ahler potential.}

\begin{document}
\hypersetup{pageanchor=false}
\makeatletter
\let\old@fpheader\@fpheader
\preprint{IFT-UAM/CSIC-26-145,
CERN-TH-2026-259}

\makeatother

\maketitle

\hypersetup{
    pdftitle={EFT strings and membranes with fluxes},
    pdfauthor={Veronica Collazuol, Alessandra Grieco, Irene Valenzuela},
    pdfsubject={EFT strings and membranes flux}
}

\newcommand{\remove}[1]{\textcolor{red}{\sout{#1}}}

\newpage

\section{Introduction \&  Conclusions} \label{sec:introduction}

Perturbative regimes of string compactifications provide a privileged setting in which to investigate universal quantum gravity consistency constraints. In four-dimensional $\mathcal{N}=1$ theories, such regimes exhibit approximate axionic shift symmetries, which severely constrain the asymptotic form of the Kähler potential and the flux-induced superpotential. The characteristic log-structure of the Kähler potential arising in string compactifications places these perturbative regimes at infinite distance in moduli space, and therefore, sensitive to the effects from the Distance Conjecture \cite{Ooguri:2006in}: a central quantum gravity constraint in the Swampland program \cite{Vafa:2005ui,Brennan:2017rbf,Palti:2019pca,vanBeest:2021lhn,Grana:2021zvf,Harlow:2022gzl,Agmon:2022thq}. According to the Distance Conjecture, an infinite tower of states becomes exponentially light in Planck units with the proper field distance 
$
m_{\rm tower}\sim e^{-\alpha\,\Delta\phi}
$. The appearance of this tower lowers the cut-off where quantum gravity effects kick in and eventually implies a drastic breakdown of the EFT. 
Understanding how this ultraviolet information is encoded in the asymptotic data of the four-dimensional EFTs is therefore central both to the Swampland program and to phenomenological applications of string compactifications.

The aforementioned axionic shift symmetries also provide a complementary probe of these perturbative/infinite distance limits given by the physics of EFT strings and EFT membranes \cite{Lanza:2020qmt}. These are extended $\frac12$-BPS objects whose cores cannot be resolved within the effective theory and therefore should be regarded as fundamental from its viewpoint. The EFT strings are magnetically charged under the axionic shift symmetries, and the scalar profile induced by their backreaction in space (denoted as the string flow) can be mapped to a particular infinite distance trajectory in moduli space. Once we have a non-vanishing superpotential, consistency of the discrete gauged axionic shift symmetries implies the existence of EFT membranes, to which the EFT string must be attached. For a flux-induced superpotential, these EFT membranes are magnetically charged under the internal fluxes. Once the existence of these objects is assumed by symmetry arguments, supersymmetry makes their tensions and charges remarkably accessible from the bottom up, as they can be derived directly from the Kähler potential and superpotential. Thus, the asymptotic structure of the EFT low energy action can be encoded in the scaling and properties of these UV extended objects.

EFT strings and membranes were originally studied in \cite{Lanza:2020qmt}, with a review  given in \cite{Lanza:2022zyg}. The relation between EFT strings and the towers of particles predicted by the Distance Conjecture was analysed in \cite{Lanza:2021udy}. A remarkable outcome of these works is that the scaling vectors\footnote{The scaling vectors are defined in terms of the variations of the masses/tensions in moduli space as
\beq \label{eq:ctmratio}
\vec\zeta= - \vec{\nabla}_{\phi} \log m(\phi) \, , \, \, \text{or} \, \, \vec\zeta = - \vec{\nabla}_{\phi} \log T(\phi) \, \,  
\eeq} of the relevant particle towers are organized in a lattice generated by those of the elementary EFT strings, since the asymptotic scaling of the mass $m_{\t}$ of the tower of states behaves as $m_{\t}^2\sim T_{\s}^\omega$ in terms of the string tension $T_{\s}$  in Planck units along the corresponding string flow. 
This relation has been observed in a large number of string compactifications \cite{Grieco:2025bjy} and has been studied in relation to various swampland constraints \cite{Martucci:2022krl, Martucci:2024trp,Marchesano:2022axe, Marchesano:2023thx, Hassfeld:2025uoy, Monnee:2025ynn, Kaufmann:2026mha, Kaufmann:2026tsy,Grieco:2026yip,  Aoufia:2026ztl, Aoufia:2026mqb, Etheredge:2026jun}, but a bottom-up rationale is still missing. The existence of these EFT strings has also been used to put constraints on different properties of the EFT, their asymptotic structure and even rule out Kähler potentials \cite{Grieco:2026yip}. 
On a different note, EFT membranes are very useful to study the asymptotic structure of the scalar potential \cite{Lanza:2019xxg,Lanza:2020qmt}, and to answer questions of phenomenological interest like whether it admits asymptotic cosmic acceleration \cite{Calderon-Infante:2022nxb}. Putting this together with the towers of states allows one to uncover potential issues of EFT consistency in relation to the tower of states \cite{Casas:2024oak}.\\

However, until now, the three types of objects—towers of particles, EFT strings, and EFT membranes—have not been incorporated simultaneously into a unified analysis of their relative asymptotic scales.
The first goal of this work is to complete this picture. We will consider the three types of objects together, both from a bottom-up and top-down perspective, providing the explicit scaling in concrete infinite distance limits of string theory compactifications. From bottom-up, we show that the scaling vectors of elementary EFT membranes also belong to the lattice generated by the EFT strings,
\beq
\vec{\zeta}_{\b}
=\sum_i w_{i,\b}\,\vec{\zeta}_{s_i}
\qquad w_{i,\b}\in\mathbb{Z},,
\eeq
in direct analogy with the particle towers studied in \cite{Lanza:2020qmt,Lanza:2021udy,Grieco:2025bjy,Grieco:2026yip}. This result follows from the asymptotic form of the superpotential and the monodromy filtration associated to the axionic shift symmetries, and provides a common organizing principle for particles, strings, and membranes near perturbative infinite-distance limits. We then determine their relative exponential rates in a range of two-modulus limits arising in F-theory and Type IIA Calabi–Yau compactifications, which were originally classified in \cite{Calderon-Infante:2022nxb} (see also \cite{Casas:2024oak}). In particular, we compute the exponential mass decay rates of the leading particle towers, EFT strings, and EFT membranes. As the latter are related to the scalar potential, this allows us to tabulate the corresponding ratios, including the exponent $\lambda$ defined through $V\sim m_{\t}^{\lambda}$ along the asymptotic gradient flow trajectories. We find that $\lambda$ is always between $\frac{2}{3}$ and $\frac{14}{3}$, which is wider than the expected range $2 \leq \lambda \leq 4$ that has been considered in the literature \cite{Montero:2022prj,Casas:2024oak}, although the cases with $\lambda< 2$ turn out to be pathological. We also find that the value $\lambda=4$ (characteristic of the Dark Dimension scenario \cite{Montero:2022prj}) also appears with this classical flux potential (without considering Casimir energies), but that it always corresponds to emergent string limits in this setup, the converse not being always true. \\

We subsequently apply this framework to the study of asymptotic accelerated expansion in 4d $\mathcal{N}=1$ theories. The scalar potential asymptotically behaves like a runaway, satisfying the de Sitter conjecture \cite{Obied:2018sgi}. This is related to the physics of EFT membranes, as shown in \cite{Lanza:2020qmt}. A positive exponential runaway can support accelerated expansion only if its logarithmic slope is sufficiently small. In four dimensions, for the scaling solutions considered here (see \cite{Calderon-Infante:2022nxb}), this requires
\beq
\gamma\equiv\frac{\lVert\nabla V\rVert}{V}<\sqrt{2},.
\eeq
This condition is in tension with strong asymptotic versions of the de Sitter conjecture \cite{Rudelius:2021azq}. Moreover, one needs to ensure that the accelerated expansion takes place within a controlled four-dimensional EFT, as otherwise the results derived from that EFT cannot be trusted. In the present context, this requires not only that the tower of particles remain above the relevant infrared scales, but also that every fundamental EFT string remain heavier than the Hubble scale,
\beq
H\lesssim T_{\rm s}^{1/2}.
\eeq
If this condition fails, the string oscillator modes enter at or below the cosmological scale and a local four-dimensional EFT containing only finitely many fields is no longer adequate. This condition on the string tension is also easier to check than the equivalent requirement that the towers of states remain heavy, since string tensions can be derived directly from the Kähler potential, whereas the masses of UV towers are not always known.

The second goal of the paper is then to leverage the physics of EFT strings and membranes to determine whether asymptotic acceleration is truly possible, both from a bottom-up and a top-down perspective.
We first ask whether the general bottom-up structure of four-dimensional $\mathcal{N}=1$ theories, together with the lattice organization of EFT strings and membranes, is sufficient to exclude asymptotic acceleration. We formulate both the acceleration condition and the requirement $T_{\rm s}^{1/2}\gtrsim H$ geometrically in terms of conditions on the convex hull of the scaling vectors. In particular, we analyse potentials dominated either by a single asymptotic term or by competing terms whose convex hull produces a parametrically shallower gradient-flow direction. We find that the bottom-up constraints alone do not yield a general obstruction: the EFT-string lattice admits a priori configurations in which the potential is sufficiently shallow for accelerated expansion while all elementary EFT strings remain parametrically heavier than the Hubble scale. Additional top-down information is therefore necessary.

We then revisit the explicit flux compactifications studied in \cite{Calderon-Infante:2022nxb}, where two candidate asymptotically accelerating trajectories were identified in the complex-structure moduli space of F-theory on elliptically fibered Calabi–Yau fourfolds with RR and NS fluxes. These candidates arise from competing terms in the flux potential and appear to evade the expected asymptotic bounds on its slope. More concretely, although each individual flux term would give rise to an excessively steep trajectory, their combination produces a gradient-flow trajectory in a runaway potential toward infinite field distance with slope $\gamma<\sqrt{2}$. These top-down constructions do not yet constitute fully fledged string-theory examples of accelerated expansion, since the trajectories involve only the complex-structure moduli and neglect the Kähler moduli. Nevertheless, they pose an interesting puzzle: if the Kähler moduli could somehow be stabilized while the complex-structure moduli continued to run, these constructions would yield the first examples of long-lived accelerated expansion at parametrically late times in string theory. What, then, prevents this scenario? In this paper, we close this loophole. Once the EFT strings are included, we find that neither candidate defines a controlled accelerating solution. Along each problematic trajectory yielding accelerated expansion, at least one EFT string becomes as light as, or lighter than, the Hubble scale. The original four-dimensional EFT therefore breaks down before the putative accelerating regime can be reached. This closes the loophole identified in \cite{Calderon-Infante:2022nxb} within the class of compactifications under consideration, leaving no remaining top-down candidate for asymptotic acceleration in this setup.  It would be interesting to investigate whether the EFT breakdown in these limits can be physically related to the criteria in \cite{Etheredge:2026cha} on the potential terms and their transformation under Weyl rescalings that imply the dS conjecture, but which are violated in the aforementioned top-down scenarios at classical level. 

The appearance of these light EFT strings also indicates that the corresponding limits within the pure complex-structure moduli space are intrinsically problematic and must receive important corrections. In other words, the leading forms of the Kähler potential and superpotential derived from these F-theory compactifications are not trustworthy in these infinite-distance limits of the complex structure moduli space. This conclusion dovetails with the quantum obstructions to $\mathcal{N}=1$ infinite-distance limits recently proposed in \cite{Kaufmann:2026fli,Kaufmann:2026mha,Kaufmann:2026tsy}. Those works show that quantum effects can obstruct an infinite-distance limit confined to the complex-structure sector and force a correlated evolution of the Kähler moduli, thereby modifying the asymptotic Kähler potential. We investigate whether the proposed corrections to the Kähler potential can restore the proper hierarchy between the EFT-string and Hubble scales in the accelerating candidates. We find that these corrections can indeed solve the hierarchy issue, since they either make the string heavier or invalidate the interpretation of such string as a valid EFT BPS string. Moreover, they also prevent the asymptotic acceleration driven by the flux potential, unless one could find a way to stabilize the quantum corrected volumes while still moving on the complex structure/dilaton directions. If such was the case, only further corrections to the flux superpotential could save the day.\\

The paper is organized as follows. In Section \ref{sec:eftstrings} we review EFT strings and membranes in four-dimensional $\mathcal{N}=1$ theories, establish the lattice structure of the membrane scaling vectors, and discuss their relation to particle towers and the asymptotic potential. In Section \ref{sec:lightstrings} we formulate the conditions for accelerated expansion and EFT validity in terms of the scaling vectors for EFT strings and membranes and perform a bottom-up two-modulus analysis. Section \ref{sec:topdown} contains the top-down study of F-theory and Type IIA Calabi-Yau compactifications, including the complete comparison between the scalings of particle towers, EFT strings, and EFT membranes. We then apply these results to the candidate accelerating trajectories of \cite{Calderon-Infante:2022nxb} and discuss their relation to the quantum obstructions of \cite{Kaufmann:2026fli,Kaufmann:2026mha,Kaufmann:2026tsy}.

\section{EFT strings and membranes} \label{sec:eftstrings}

\subsection{Review} 
\label{sec:review}

EFT strings and membranes are low-codimension objects ubiquitous to EFTs and particularly relevant in 4d supersymmetric theories with approximate axionic shift symmetries and F-term potentials. The analysis of their properties and the consequences they yield on Swampland conjectures, which we review briefly in the following, has been first systematically carried out in \cite{Lanza:2020qmt,Lanza:2021udy, Lanza:2022zyg}. These objects are regarded as \textit{fundamental} in a 4-dimensional effective theory, implying that their cores cannot be resolved within the EFT itself. This means that their tension is bounded from below by the EFT cut-off $\Lambda$, i.e.
\begin{equation} \label{eq:fundamental}
    \Lambda^2<T_{\s} < M_{\text{P}}^2 \, , \qquad \text{and} \qquad \Lambda^3<T_{\b} < M_{\text{P}}^2 \Lambda \, ,
\end{equation}
for strings and membranes, respectively.
In 4d $\mathcal{N}=1$ theories, EFT strings and membranes are BPS objects, thus their tension is completely determined once the Kähler potential $K$ and the superpotential $W$ are known. 
Recall the bosonic action for a set of chiral multiplets $\{\phi^i\}$ in these theories, 
\begin{equation}\label{eq:4daction}
    S = \int \left(\frac{M_{\text{Pl}}^2}{2}R \star 1 -M_{\text{Pl}}^2K_{i\bar{j}}\text{d}\phi^i \wedge \text{d}\phi^{\bar{j}} - V \star 1\right).
\end{equation}
Here $V$ is an F-term scalar potential coming from the superpotential $W$, turned on by the presence of fluxes $f_A \in \mathbb{Z}$ in such a way that  
\begin{equation} \label{eq:scalarpotentialn1}
    V =e^{K} (K^{I\bar{J}}D_{I}W\overline{D}_{\bar{J}}\overline{W} - 3 |W|^2) \, .
\end{equation}

In regimes where the Kähler potential is approximately shift-symmetric, we can resort to a dual formulation for both the effective action and the fluxes. In particular, consider a subset $\{t^a\}$ of the chiral fields $\{\phi^i\}=\{t^a,\chi^\alpha\}$, periodic in their real component $a^a=\text{Re} \ t^a \simeq \text{Re} \ t^a + e^a, \ e^a \in \mathbb{Z}$.  Asymptotically, these periodic directions are promoted to axionic shift symmetries,
\begin{equation}
    \text{Re} \ t^a\to \text{Re} \ t^a +\xi e^a, \ \xi \in \mathbb{R}.
\end{equation}
In this asymptotic regime, the axionic shift symmetries are made manifest by the Kähler potential $K$ depending only on the saxions $s^a = \text{Im} \ t^a$ at perturbative level.

Hence, these are perturbative limits: the leading non-perturbative corrections that would break the shift symmetry can arise from $\frac{1}{2}$-BPS instantons, giving a contribution to the Kähler potential of $\mathcal{O}(e^{2i\pi m_at^a})$ in the instanton charges $m_a \in \mathbb{Z}$. In asymptotic limits where  linear combinations of $\{s^a\}$ become very large, these corrections are exponentially suppressed. Thus, the leading order of the Kähler potential in these regimes is given by
\begin{equation}\label{eq:kahler-pot}
    K= -\log \left[P(s)+
    \mathcal{O}(e^{-2\pi m_as^a})\right],
\end{equation}
where $P(s)$ is a homogeneous function of positive integral degree in the saxions. Away from this asymptotic regime, the breaking of shift symmetry is measured by a set of instanton charges $\mathcal{C}_I$. Hence, we can define each perturbative regime by requiring that 
\begin{equation}\label{eq:pert-reg}
\braket{\mathbf{m},\mathbf{s}}= m_is^i \gg 1 \quad \forall \mathbf{m} \in \mathcal{C}_I,
\end{equation}
where $\mathbf{s}$ denotes all the saxions that become large.
Then the perturbative region is given by the saxionic cone $\Delta$ in its \textit{deep interior},
\begin{equation}
    \Delta = \left\{s^i\in N_\mathbb{R} \ \forall | \ \braket{\mathbf{m},\mathbf{s}}>0,\forall \mathbf{m} \in \mathcal{C}_I \right\},
\end{equation}
where $N_\mathbb{R}$ is a vector space of appropriate dimension.

Thanks to the asymptotic shift symmetry of the Kähler potential, it is possible to dualize each chiral field into a \textit{dual saxion} $\ell_a$ and a two-form potential $\mathcal{B}_{2a}$ given by
\begin{equation}
     \ell_a = -  \frac{1}{2}\frac{\partial K}{\partial s^a}, \quad \mathcal{H}_{3a} = \text{d}\mathcal{B}_{2a} = -M^2_{\text{P}}G_{ab}  \star\text{Re} \ t^a,
\end{equation}
where $G_{ab}$ is the shift-symmetric field space metric,
\begin{equation}\label{eq:mspace-metric}
    G_{ab} = \frac{\partial^2 K}{\partial s^a \partial s^b}.
\end{equation}
Computed the dualized version of \eqref{eq:4daction}, we can add to it the action of a $1/2$-BPS 4d string, magnetically coupled with charge $e^a$ to the two-form $\mathcal{B}_{2a}$. Then the tension and physical charge of such string are simply given by 
\begin{equation}
    \mathcal{T}_{\textbf{e}} = M^2_{\text{P}} |e^a\ell_a|, \quad \mathcal{Q}_{\textbf{e}} =  M_{\text{P}}\sqrt{G_{ab}e^ae^b},
\end{equation}
where $\mathbf{e}=(e^1,\dots,e^n)$ and $a,b= 1,\dots,n$.
 As we approach its core, the string backreacts on the space-time geometry and saxions acquire a non-trivial profile given by
\begin{equation}\label{eq:s-flow}
    s^i = s^i_0 + e^i\sigma, \quad \sigma= -\frac{1}{2}\log\left(\frac{r}{r_0}\right),
\end{equation}
where the string core is located at $r=0$. At the same time, the axions $a^a=Re \ t^a$ undergo a monodromy when performing a rotation around the core,
\begin{equation}
    a^i =a^i_0\frac{\theta}{2\pi}e^i.
\end{equation}
From \eqref{eq:s-flow}, we can see that close to the string core the backreaction profile of the saxions diverges, as $s^i \to e^i \cdot \infty$ as $r\to 0$, while $a^i$ remains constant.  This profile \eqref{eq:s-flow} maps to a path  in moduli space which is known as \textit{string flow}.

\textit{EFT strings} are those 1/2-BPS strings whose string flow drives us deeper into the perturbative regime, so the non-perturbative corrections remain subleading along the flow as approaching  the string core.
The logarithmic form of the K\"ahler potential implies  that the string flow of EFT strings bring us to an infinite distance boundary of moduli space.

In general, following the definition of the perturbative regime as in \eqref{eq:pert-reg}, we can identify its associated set of EFT string charges with a discretization of the saxionic cone $\Delta$,
\begin{equation}
    \mathcal{C}_\text{S}^{\text{EFT}} = \overline{\Delta}\ \cap\ N_\R,
\end{equation}
where $\overline{\Delta}$ is the closure of $\Delta$. 
Strings that have only one non-vanishing unit charge are called \textit{elementary} EFT strings,

\begin{equation}\label{eq:elementary-string}
\text{Elementary EFT string}\rightarrow \mathcal{T}_{a}= M_{\text{P}}^2\ell_a.
\end{equation}

We have seen that saxions can be dualized into dual saxions which in turn control the tension of EFT strings, magnetically charged under the dual two-forms. Analogously, it is possible to dualize the flux quanta $f_A$ in terms of three-form potentials $A_3$ to which 4d membranes can couple. This is a non-trivial procedure, as explained in \cite{Farakos:2017jme,Bandos:2018gjp}: in particular, it has been shown \cite{Lanza:2019xxg} that not all fluxes can be dualized in three-forms from the low energy EFT perspective, mainly because the EFT UV cut-off $\Lambda$ imposes constraints on which membranes can mediate transitions described within the EFT, as can be read from \eqref{eq:fundamental}. This subset of 1/2-BPS membranes organize in a lattice; once this is established, we can define the lattice of \textit{dynamical fluxes} as 
\begin{equation}
    \Gamma_{\text{EFT}} = \left\{ \text{Fluxes dualizable to 3-forms and consistent with} \ \frac{T_b}{M_{\text{P}}^2} <\Lambda\right\}.
\end{equation}

In this dual framework, one can add to \eqref{eq:4daction} the action of a 4d membrane with magnetic charge $q_A \in \Gamma_{\text{EFT}}$ under the dual three-forms. This membrane implements a flux jump $f_A \to f_A + q_A$ at the level of the EFT, such that the scalar potential will be different on each side of the membrane. From the dualized action, we can read off the tension and physical charge of these \textit{EFT membranes} as
\begin{equation}
    T_\mathbf{q} = 2 e^{\frac{K}{2}} |W_\mathbf{q}|, \quad \mathcal{Q}_\mathbf{q}^2= 2V_\mathbf{q}
\end{equation}
where the superpotential $W_\mathbf{q}$ is turned on by the set of fluxes $\mathbf{q}= (q_1,\dots, q_m)$ and $A=1,\dots,m$. Also, $V_\mathbf{q}$ can be read off from \eqref{eq:scalarpotentialn1}. 

In terms of \textit{period vectors} $\Pi^A(\phi)$ we can also write
 \begin{equation} \label{eq:superpotential}
   W_{\mathbf{q}}(\phi) = M^3_{\text{P}}q_A\Pi^A(\phi),
 \end{equation}
where we are not considering possible contributions from non-dynamical fluxes and non-perturbative effects.

\subsection{EFT strings lattice}

\subsubsection*{Towers of particles in the lattice}

\begin{figure}[t]
    \centering
    \includegraphics[width=0.6\linewidth]{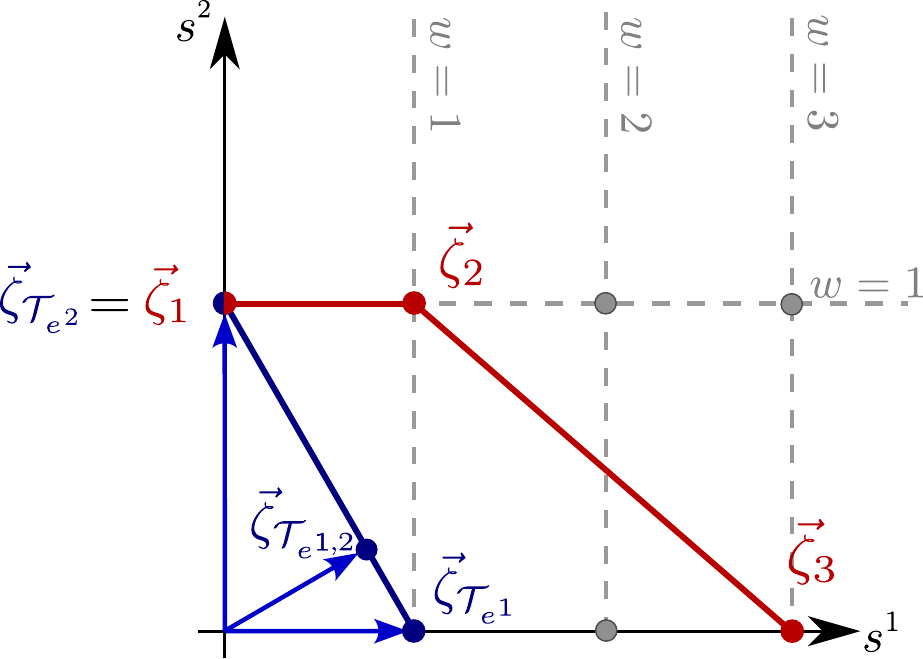}
\caption{Example of convex hull of scaling vectors $\vec{\zeta}_i$ (in red) corresponding to towers of states becoming light in a given growth sector. In blue, scaling vector corresponding to the EFT strings generating the lattice in which light towers live. Notice that not all points of the lattice are populated: in grey, the points of the lattice that would correspond, if occupied, to towers decompactifying a non-integer number of dimensions.
\label{fig:chullex}}
\end{figure}

A Kähler potential of the form \eqref{eq:kahler-pot} implies that the tension of EFT strings, depending only on the dual saxions, vanishes asymptotically along the string flow \eqref{eq:s-flow},
\begin{equation}
\mathcal{T}_\textbf{e}\sim\frac{\Mpf^2}{\sigma}\to 0 \quad \text{with}\quad \sigma \to \infty.
\end{equation}
One expects then to find such EFT strings becoming light in Planck units at every infinite distance limit of moduli space. In \cite{Lanza:2021udy}, a large number of 4d \n string compactifications has been surveyed to identify the microscopic origin of EFT strings. In the same work, an \textbf{Integral Scaling Relation} between the tension of EFT strings and leading towers of states becoming light was found,
\begin{equation}\label{eq:integral-scaling}
\begin{split}
    m^2_\star\simeq  \Mpf^2\left(\frac{\mathcal{T}_{\mathbf{e}}}{\Mpf^2}\right)^w &\text{with} \ w=1,2,3 \\ \text{as} \ s^i= s^i_0+e^i\sigma \ &\text{and} \ \sigma \to \infty,
\end{split}
\end{equation}
where $m_\star$ is the mass scale of the leading tower and $w$ is known as \textit{scaling weight}. This relation was further tested in \cite{Grieco:2025bjy}, where it was found to be valid not only for the leading towers but also for subleading towers not given by bound states of other towers. To see this better, \eqref{eq:integral-scaling} can be recast as a local condition in moduli space by defining the following \textit{scaling vectors}
\begin{equation}
\label{eq:scaling-vec}    \vec{\zeta}_i=-\vec{\nabla}\log \frac{m_i(\vec{\varphi})}{\Mpf}, \quad \ \vec{\zeta}_\mathcal{T}=-\frac{1}{2}\vec{\nabla}\log \frac{\mathcal{T}_\mathbf{e}(\vec{\varphi})}{\Mpf^2},
\end{equation}
with derivatives in the moduli. Thus we can rewrite \eqref{eq:integral-scaling} as
\begin{equation}\label{eq:integer-scaling-loc}
\vec{\zeta}_\star \cdot \vec{\zeta}_\mathcal{T} =w |\vec{\zeta}_\mathcal{T}|^2, \quad \text{with} \ w=1,2,3.
\end{equation}
Here, the scalar product is taken with respect to the moduli space metric \eqref{eq:mspace-metric}. In \cite{Grieco:2025bjy} it was observed, by scanning various 4d \n compactifications, that this relation is valid for any EFT string and any subleading tower being a vertex of the towers' \textit{convex hull}. Interestingly, restricting our set of EFT string only to elementary ones is enough to characterize all the infinite distance limits in a given perturbative region. This is because \eqref{eq:integer-scaling-loc} effectively defines a \textit{lattice condition}\footnote{For the sake of precision, in a generic growth sector this is not a full lattice (we will not have any $w \in \mathbb{Z}$) but a restriction of it.} for the scaling vectors in which elementary EFT strings play the role of lattice generators. Hence in each perturbative limit the light towers of states will occupy a lattice site in the $\zeta$-space, although it is not necessary for each site to be populated.  A qualitative example of this is shown in Figure \ref{fig:chullex}.
 
The presence of empty lattice sites is explained by the inconsistency of the putative tower that would occupy them: for example, in Figure \ref{fig:chullex}, gray points are at such a distance from the origin that a light tower there would have an exponential rate incompatible with the Emergent String Conjecture (ESC) \cite{Etheredge:2022opl,Etheredge:2024tok}. In particular, gray points would represent towers decompactifying a non-integer number of dimensions. These observations in 4d \n string compactifications are the starting point of \cite{Grieco:2026yip}, where the logic is reversed and these top-down patterns are taken as assumptions from the bottom-up point of view. This results in a finite classification of 4d \n convex hulls of light towers in each growth sector, in terms of the lattice generated by EFT strings and compatibility with the ESC.

\subsubsection*{EFT membranes in the lattice}
In this subsection, our goal is to show that the $\zeta-$ vectors of elementary EFT membranes also belong to the lattice generated by the fundamental EFT strings, as introduced in \cite{Grieco:2025bjy}. 

\begin{figure}[t]
    \centering
    \includegraphics[width=0.7\linewidth]{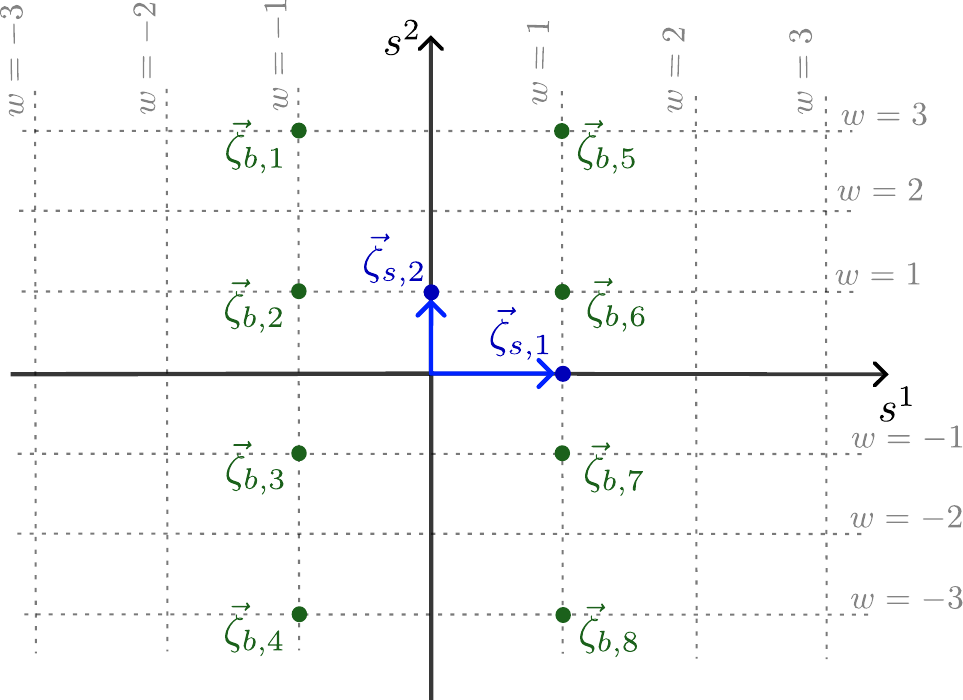}
\caption{Example of lattice on which of branes' scaling vectors $\vec{\zeta}_{b,i}$ live. In blue, scaling vectors corresponding to the EFT strings generating the lattice in which both branes and light towers live. Notice that not all points of the lattice are populated: this is what we will observe in top-down examples. Notice also that the lattice is the same one for particles scaling vectors, extended also to negative scaling weights. 
\label{fig:lattice-ex}}
\end{figure}

Let us consider a limit where the $I$ saxions are taken to infinity, i.e. $s^i \to \infty$ for $i = 1, \dots, I$, in a given growth sector where the moduli scale as
\begin{equation}
    \mathcal{R}_{12\dots I} = \left\{ \varphi^i = a^i + is^i \Big{|}  \frac{s^1}{s^2} > \delta, \dots , \frac{s^{I-1}}{s^I} > \delta, s^I > \delta,\, \ a^i < \beta \right\} \, ,
\end{equation}
and the leading contribution to the Kähler potential takes the form\footnote{In the language of limiting Mixed Hodge Structure, $n_i = n'_i - n'_{i-1}$, with $n_0 = 0 $.  Hence, $n'_i = \sum_{j=1}^i n_i$.}
\begin{equation}
    K = - \log \left( s_1^{\, n_1} s_2^{\, n_2} \dots s_I^{\, n_I} \right)  \, .
\end{equation}
As worked out in \cite{Grimm:2018ohb}, from the Nilpotent Orbit Theorem \cite{Schmid1973}, near each divisor $\mathcal{D}_i = \{e^{2 \pi i \varphi^i} =0 \}$ (with all the other moduli $\chi$ fixed) the periods appearing in the superpotential \eqref{eq:superpotential} can be written as
\begin{equation}
    \Pi(\varphi, \chi) = e^{\varphi^i N_i^{T}} \Pi_0(\chi) + \mathcal{O}(e^{2\pi i \varphi}) \, ,
\end{equation}
where $\Pi_0$ is a period independent of the asymptotic fields, and $N_i$ are nilpotent generators of nilpotency order $d_i \in \mathbb{N}$, i.e. 
\begin{equation}
    (N_i^{T})^{d_i} \Pi_0(\chi) \ne 0 \, , \qquad (N_i^{T})^{d_i+1} \Pi_0(\chi) = 0 \, .
\end{equation}
In what follows, we will be interested in the perturbative part of the superpotential and will neglect the exponentially suppressed contribution, such that the superpotential turns out to be a finite polynomial in the saxions. This in turn allows to write the tension of a membrane with charge $\mathbf{q}$ as  
\begin{equation} \label{eq:tension}
    T_\mathbf{q} = 2 e^{\frac{K}{2}} \Big{|}\sum_{k=0}^{d/2} \frac{1}{k!} (is)^k N^k \bold{\rho}(\mathbf{q},a) \Pi_0\Big{|} \, ,
\end{equation} 
where we defined $\bold{\rho}(\mathbf{q},a) = e^{aN} \mathbf{q}$.  We restricted for simplicity to the expression for a single modulus, which can be straightforwardly generalised. Using the monodromy filtration that the nilpotent operator induces on the flux lattice in pairwise orthogonal spaces $\Gamma_{\mathbf{r}}$,
\begin{equation} \label{eq:filtration}
    \Gamma_{\text{F}} = \bigoplus_{\mathbf{r}} \Gamma_{\mathbf{r}} \, , \quad \text{with} \quad \mathbf{r} = (r_1, \dots, r_I) \, , \, \,  - d_i \leq r_i \leq d_i \, ,
\end{equation}
it can be shown with the tools of limiting Mixed Hodge Structures \cite{hodgestructure} that for each asymptotic limit one can always find a basis of ``elementary flux vectors" $\mathbf{q}_{\mathbf{r}} \in \Gamma_{\mathbf{r}}$ for which the leading contribution to the tension of these membranes -- that are called \textit{elementary} membranes -- scales as 
\begin{equation} \label{eq:branes}
    T_{\text{q}_{\mathbf{r}}} \simeq T_0(\chi,\bar{\chi})\rho_{\mathbf{r}}(\mathbf{q},a) s_1^{\tilde{r}_1/2}s_2^{\tilde{r}_2/2} \dots s_k^{\tilde{r}_k/2} \, ,
\end{equation}
where $\tilde{r}_i=r_i-r_{i-1} +d_i - d_{i-1} - n_i \in \mathbb{Z}$\footnote{ Remember that in the language of Mixed Hodge structure, $d_i \leq n'_i$ and $-n'_i \leq r_i - n'_i+d_i \leq 2d_i -n'_i$. Hence, $ -2 d_{i-1}  -n_i \leq \tilde{r}_i \leq 2d_i - n_i$.  \label{footnote4}}, and in general $k\leq I$, such that
\begin{equation}
    \frac{\partial T_{\text{q}_{\bold{r}}}}{\partial s^i} = - \frac{1}{2} \frac{\tilde{r}_i}{n_i} \partial_i K \, T_{\text{q}_{\bold{r}}} \, . 
\end{equation}
Hence, in flat coordinates, eq. \eqref{eq:scaling-vec} reads
\begin{equation}\label{eq:brane-zvec}
    \left(\zeta_{\text{q}_{ \mathbf{r}}}\right)_a = - \frac{\tilde{r}_i}{\sqrt{2 n_i}} \delta_{ai} \, .
\end{equation}
On the other hand, the $\zeta$-vector for elementary EFT strings associated with the saxionic flow of $s^i$, $\vec{\zeta}_{s^i}$\footnote{Here and henceforth we will use this notation for the scaling vector of EFT strings.}, can be computed as 
\begin{equation} \label{eq:zetastr}
    \left( \zeta_{\s_i} \right)_a = \sqrt{\frac{1}{2 n_i}} \delta_{ai} \, .
\end{equation}
Equations \eqref{eq:brane-zvec} and \eqref{eq:zetastr} can be combined to obtain 
\begin{equation}
    \frac{\vec{\zeta}_{\text{q}_{\bold{r}}}\cdot \vec{\zeta}_{\s_i}}{|\vec{\zeta}_{\s_i}|^2} = -\tilde{r}_i \in \mathbb{Z} \, .
\end{equation}
which defines a lattice, generated by elementary EFT strings, for the branes scaling vectors (see also \cite{Etheredge:2026cha}). Remarkably, this is the same lattice as the one that supports the convex hull of light towers \eqref{eq:integer-scaling-loc}. See Figure \ref{fig:lattice-ex} for an example of this. As we will see in section \ref{sec:topdown}, in all the type II/F-theory top-down examples that we explore we always have $\tilde{r}_i\in[-3,3]$.

\subsection{Bounds on the potential rate}\label{sec:bounds-pot}

 Restricting to the gravitational and moduli sector, the effective description of the \n\, theories reviewed in section \ref{sec:review} are of the type
\begin{equation}
\label{eqn:eftaction}
    S = \int d^4 x \sqrt{-g} \left( \frac{R}{2} + \frac{1}{2} g^{\mu \nu} G_{ij} \partial_{\mu} \varphi^i \partial_{\nu} \varphi^j - V(\varphi) \right)
\end{equation}
in 4 dimensional Planck units, where the scalar fields $\varphi^i= a^i + i s^i$ group in chiral multiplets. If we assume that the (moduli-dependent) potential is positive, it make sense to define the {\it dS coefficient} 
\begin{equation} \label{eq:dscoefficient}
    \gamma = \frac{||\nabla V(\varphi)||}{V(\varphi)} \, ,
\end{equation}
where the Laplacian is meant to be the one in moduli space. 

Along its gradient flow, we can parametrise the scaling of the potential with the mass of the lightest tower as
\begin{equation} \label{eq:lambdaV}
    V \sim m_{\t}^{\lambda} \, ,
\end{equation}
in Planck units. If the gradient flow aligns with some asymptotic geodesic in moduli space, then $m_{\t} \sim e^{-\alpha \, \Delta\phi }$ in terms of the geodesic field distance in moduli space $\Delta\phi \equiv \int d \tau \sqrt{G_{ij} \frac{\partial \varphi^i}{\partial \tau}\frac{\partial \varphi^j}{\partial \tau}}$. In general, we can define $\alpha$ as the projection of the scaling vector \eqref{eq:integral-scaling} over that particular trajectory with unit vector $\hat{t}$
\begin{equation}
    \alpha = \vec{\zeta}_{\t} \cdot \hat{t} \, .
\end{equation}
Hence, it is clear that asymptotically $\lambda=\frac{\gamma}{\alpha}$. 
From \cite{Montero:2022prj} -- and bottom-up arguments given in \cite{Casas:2024oak} (see also \cite{Rudelius:2022gbz}) -- it is expected that
\begin{equation}\label{eq:lambdabound}
    2\leq \lambda \leq 4 \, \leftrightarrow \, 2 \alpha\leq \gamma \leq 4 \alpha \, .
\end{equation}
The lower bound $\lambda \geq 2$ can be motivated by the Higuchi bound \cite{Higuchi:1986py}, assuming that it still applies to quasi-dS time-dependent solutions.

Let us first focus on the case in which the potential is dominated by a single flux term $V \sim T_{\b}^2$,  so that the scaling of the associated brane tension behaves as \begin{equation} \label{eq:lambda}
    T_{\b} \sim m_{\t}^{\lambda/2} \, .
\end{equation}
In that case, the bound \eqref{eq:lambdabound} directly applies to the brane tension scaling, which also needs to be compatible with \eqref{eq:fundamental}. We can now take the EFT cut-off in \eqref{eq:fundamental} to be the species scale $\Lambda_{\text{sp}}$ \cite{Dvali:2007wp}, as this is the maximum value it can reach. 
Whenever we have only one light tower of states, the scaling of the species scale $\Lambda_\text{sp}$ is related to the scaling $\alpha$ of the tower as \cite{Castellano:2021mmx}
\begin{equation}\label{eq:spscale}
    \Lambda_\text{sp} = m_{\t}^{\frac{n}{n+2}}=e^{-\sqrt{\frac{n}{2(n+2)}} \Delta \phi} \, ,
\end{equation}
along the gradient flow trajectory of the mass of the tower. 
Here we have used the fact that the exponential rate of the leading tower in this infinite distance limit is\footnote{This expression for $\alpha$ is valid for a limit in which warping effects are diluted; see \cite{Raucci:2026fzp} for more discussion on this.} (in $d=4$) $\alpha= |\vec{\zeta}_t|=\sqrt{\frac{n+2}{2n}}$, where $n$ is the number of decompactifying dimensions and formally $n\to \infty$ for emergent strings.

Assuming that the gradient flow for the potential and the tower align, then the bounds \eqref{eq:fundamental}, which in ($d=4$) Planck units are given by
\begin{equation}\label{eq:brane-species-bounds}
     \Lambda_\text{sp}^3<T_{\b} < \Lambda_\text{sp} \, ,
\end{equation} 
can be re-written as
\begin{equation}\label{eq:lambda-species-bound}
   \frac{2n}{2+n}< \lambda< \frac{6n}{2+n} \, .
\end{equation}
In particular, for a decompactification limit with $n=1$ we have $2/3<\lambda<2$ and for a string limit ($n\to \infty$) $2<\lambda<6$. For an $n=1$ tower, this seems incompatible with the bound \eqref{eq:lambdabound}. 

However, notice that the relations \eqref{eq:lambda}-\eqref{eq:lambda-species-bound}, as stated above, are specialized to  limits where the species scale, the leading tower and the brane tension are perfectly aligned.\footnote{This will happen when we only have one light tower of states, and one single flux term dominating the scalar potential, such that the gradients of the tower mass and the potential point in the same direction of moduli space. This is an extremely fine-tuned situation that it is probably only realized in one-dimensional moduli spaces.} This is equivalent to say that their scaling vectors computed as in \eqref{eq:scaling-vec} are proportional to each other. In general, however, none of these three scaling vectors would be aligned. To reformulate our statement in the more general case it is necessary to pick a direction $\hat{u}$ towards infinite distance and then define
\begin{equation}\label{eq:eff-coeff}
    \alpha_\text{eff}= \vec{\zeta}_t\cdot \hat{u}; \quad z_\text{eff}= \vec{Z}_\text{sp}\cdot \hat{u}; \quad \gamma_\text{eff}=\vec{\gamma}\cdot \hat{u}; \quad \lambda(\hat{u})= \frac{\gamma_\text{eff}}{\alpha_\text{eff}}
\end{equation}
where $\vec{\zeta}_t$, $\vec{Z}_\text{sp}$, and $\vec{\gamma}$ are the scaling vectors \eqref{eq:scaling-vec} of the leading tower, species scale, and (a single) leading scalar potential term, respectively. Then \eqref{eq:brane-species-bounds} becomes equivalent to imposing
\begin{equation}
   2\frac{z_\text{eff}}{\alpha_\text{eff}}< \lambda(\hat{u})<6\frac{z_\text{eff}}{\alpha_\text{eff}}.
\end{equation}
If the species scale and leading tower are aligned, then the above inequality reduces to the form in \eqref{eq:lambda-species-bound} with the difference that $\lambda=\lambda(\hat{u})$ depends on the specific moduli space direction. In fact, the pattern $\vec{\zeta}_t\cdot \vec{Z}_\text{sp}=1/2$ \cite{Castellano:2023jjt,Castellano:2023stg} reduces to $|\vec{Z}_\text{sp}|= 1/2|\vec{\zeta}_t|$ for the vectors aligned, so $z_\text{eff}=\alpha_\text{eff}/2|\vec{\zeta}_t|^2$. This also occurs if $\hat{u}=\vec{\zeta}_t/|\vec{\zeta}_t|$, since in this case the pattern fixes $z_\text{eff}= 1/2|\vec{\zeta}_t|$.

In Section \ref{sec:lightstrings}, we will reevaluate these bounds taking into account the lattice structure of the scaling vectors for the branes, including the cases in which the potential is dominated by several flux terms. Moreover, we will compute different values of $\lambda$ from top-down examples in Section \ref{sec:topdown}. For simplicity, from now on we will drop the effective subscript in \eqref{eq:eff-coeff} and compute all these quantities along the specified gradient flow trajectories of the potential.

\section{Accelerated expansion and light strings} \label{sec:lightstrings}
In this section, we first review the conditions that the asymptotic potential must satisfy to yield accelerated expansion at parametrically late times, while keeping the EFT under control. Following \cite{Calderon-Infante:2022nxb}, we explain how these condition can be visualized geometrically using the convex hull of the scaling vectors for EFT strings and membranes in \eqref{eq:gammagradientflow} and \eqref{eq:rhos}. Then we investigate whether these conditions can a priori be satisfied from a bottom-up perspective in $\mathcal{N}=1$ theories using a 2-moduli setup for illustration.

\subsection{Conditions for acceleration and validity of EFT} \label{sec:eftvalidity}

\subsubsection*{Condition for having an accelerated expansion}

We focus on asymptotic field trajectories where the potential goes to zero exponentially, as generically happens in string theory compactifications. In general, in systems as \eqref{eqn:eftaction} with scalars in the presence of a scalar potential, there can be different attractor solutions, but we will focus on the simpler case of scaling solutions as in \cite{Rudelius:2022gbz,Shiu:2023fhb,Calderon-Infante:2022nxb}, that do not require non-vanishing spatial curvature nor matter. The condition of accelerated expansion translates in the dS coefficient \eqref{eq:dscoefficient} satisfying 
\begin{equation} \label{eq:accexp1}
    \gamma < \frac{2}{\sqrt{d-2}} \, .
\end{equation}
On the other hand, from the Swampland perspective, the {\it dS Conjecture} \cite{Obied:2018sgi} states that, in asymptotic regimes of a QG moduli space, the dS coefficient satisfies
\begin{equation}
\label{eq:dsconj}
    \gamma > c_d \, ,
\end{equation}
with $c_d$ a (possibly $d$-dependent) constant of order one. According to the {\it strong dS conjecture} \cite{Rudelius:2021azq}, for the strong energy condition to be satisfied asymptotically at late times 
\begin{equation}
\label{eq:strongds}
    c_d = \frac{2}{\sqrt{(d-2)}} \, .
\end{equation}
which excludes accelerated expansion in asymptotic/weak coupling regimes, as it is in tension with \eqref{eq:accexp1}.
On the other hand, according to the {\it Trans-Planckian Censorship Conjecture} (TCC) \cite{Bedroya:2019snp}, for a generic form of the potential in the asymptotic limits it must hold
\begin{equation}
\label{eq:tcc}
    c_d = \frac{2}{\sqrt{(d-1)(d-2)}} \, ,
\end{equation}
in order for Hubble scale perturbations not to come from trans-Planckian scales at early times upon cosmological expansion. Though the TCC seems weaker than the strong dS conjecture, for exponential runaways -- as arising in the theories under consideration here -- the TCC actually implies the strong dS conjecture \cite{Bedroya:2019snp}.

One can also recover the bound \eqref{eq:dsconj} with \eqref{eq:strongds} from the Sharpened Distance Conjecture \cite{Etheredge:2022opl} if one additionally imposes that the mass of the tower must remain above the Hubble scale $m_{\t}>H$ \cite {Higuchi:1986py,Rudelius:2022gbz,Montero:2022prj,Casas:2024oak}. Taking $m_{\t} \sim e^{-\alpha \, \Delta \phi}$, one gets
\begin{equation}
   \alpha\geq \frac1{\sqrt{d-2}}\to \gamma\geq \frac{2}{\sqrt{(d-2)}}
\end{equation}

These bounds on the asymptotic exponential decay rate of the scalar potential have been checked and verified in plethora of top-down string theory constructions, see for example \cite{Hertzberg:2007wc,Garg:2018zdg,ValeixoBento:2020ujr,Andriot:2020lea,Cicoli:2021fsd,Andriot:2022xjh,Calderon-Infante:2022nxb,Cremonini:2023suw,Casas:2024oak,Deffayet:2023bpo,Deffayet:2024hug}. However, a possible loophole was pointed out in \cite{Calderon-Infante:2022nxb}, which we will review in more detail below. This loophole was explicitly realised in certain \n\, string and F-theory flux compactifications  that \emph{apparently} give rise to a small dS coefficient as \eqref{eq:accexp1} in  asymptotic limits of the moduli space \cite{Calderon-Infante:2022nxb}, as we will explain in Section \ref{sec:topdown}. To address this loophole, we are going to focus now on accelerated cosmologies in general \n\, setups in the presence of a superpotential $W$ inducing a scalar potential that can be expanded as
\begin{equation} \label{eq:potentialexpansion}
    V(\varphi) = \sum_{\bold{l} \in \mathcal{E}} V_{\bold{l}} \, .
\end{equation}
 Here, the physical trajectories for the scalar fields $\varphi^i(\tau)$ at late times are of {\it gradient flow} type \cite{Calderon-Infante:2022nxb}, i.e. they satisfy the linear equation 
\begin{equation}
    \frac{d\varphi^i(\tau)}{d \tau} = - f(\tau) G^{ij} \frac{\partial V(\tau)}{\partial \varphi^j} \, ,
\end{equation}
with $f(\tau)$ a smooth positive function that can be set to 1 by reparametrization. Depending on the fluxes that are turned on, only some $\bold{l} \in \mathcal{E}$ are selected in \eqref{eq:potentialexpansion}, giving different asymptotic behaviours. As originally explained in \cite{Calderon-Infante:2022nxb}, the problem of finding these trajectories -- together with the dS coefficient $\gamma$ along them and hence the compatibility with either  \eqref{eq:accexp1} or \eqref{eq:dsconj} -- can be equivalently recast in a geometric fashion for a potential of the form \eqref{eq:potentialexpansion} by introducing the {\it dS ratio} associated with each of the terms $V_{\bold{l}}$ of the potential \cite{Calderon-Infante:2022nxb}
\begin{equation}
    \mu^a_{\bold{l}} = - \delta^{ab} E_{b}^i \frac{\partial_i V_{\bold{l}}}{V_{\bold{l}}} \, ,
\end{equation}
where the vielbein, $E_{a}^i$, is defined such that $E^i_{a} G_{ij} E^j_b = \delta_{ab}$.
The dS coefficient along the gradient flow direction in the tangent space $\hat{t}_{\text{gf}}$ (appropriately normalized to 1) can then be expressed as
\begin{equation} \label{eq:gammagradientflow}
    \gamma_{\text{gf}} = \min_{\bold{l}} \left( \hat{t}_{\text{gf}} \cdot \vec{\mu}_{\bold{l}} \right) \, 
    \end{equation}
where the minimization of the projections onto $\hat{t}_{\text{gf}}$ should be done over the set of $\vec{\mu}_{\bold{l}}$ associated with the specific potential $V(\varphi)$ at hand. In general, there can be more than one $\vec{\mu}_{\bold{l}}$, having the same projection along the gradient flow trajectory. Moreover, by defining $\mathcal{C}_{\mu}$ to be the convex hull of the $\{ \vec{\mu}_{\bold{l}}\}_{\bold{l}\in \mathcal{E}}$, it also holds that
\begin{equation}
\label{eq:gammagf}
    \gamma_{\text{gf}} = \min \left(\text{dist}(0,\mathcal{C}_{\mu}) \right) \, ,
\end{equation}
such that essentially the gradient flow direction is the one (and only one) whose tangent vector defines the minimal distance of the convex hull to the origin of the tangent space, and its length determines the fall-off $\gamma$ of the potential asymptotically.

In this formulation, the conjecture \eqref{eq:gammagf} is then verified if the convex hull $\mathcal{C}_{\mu}$ lies outside the ball of radius $c_d$ -- given either by \eqref{eq:strongds} or \eqref{eq:tcc} \cite{Calderon-Infante:2022nxb}.

In particular, as mentioned above, one can focus on two different cases, namely
\begin{itemize}
    \item (I): There is only one $\vec{\mu}_{\text{dom}}$, coinciding with one of the $\vec{\mu}_{\bold{l}}$, and with the gradient flow direction. For consistency, one must ask that such a $\vec{\mu }_{\bold{l}}$ lies in the first quadrant, so that the potential vanishes asymptotically.
    \item (II): There are two competing terms in the potential,\footnote{This situation is much richer, and in general allows more easily for accelerated expansion \cite{Calderon-Infante:2022nxb}.} defining an edge whose minimal distance point to the origin of the tangent space defines the tangent vector to the gradient flow trajectory. 
\end{itemize}

\begin{figure}[t]
    \centering
    \includegraphics[width=0.48\linewidth]{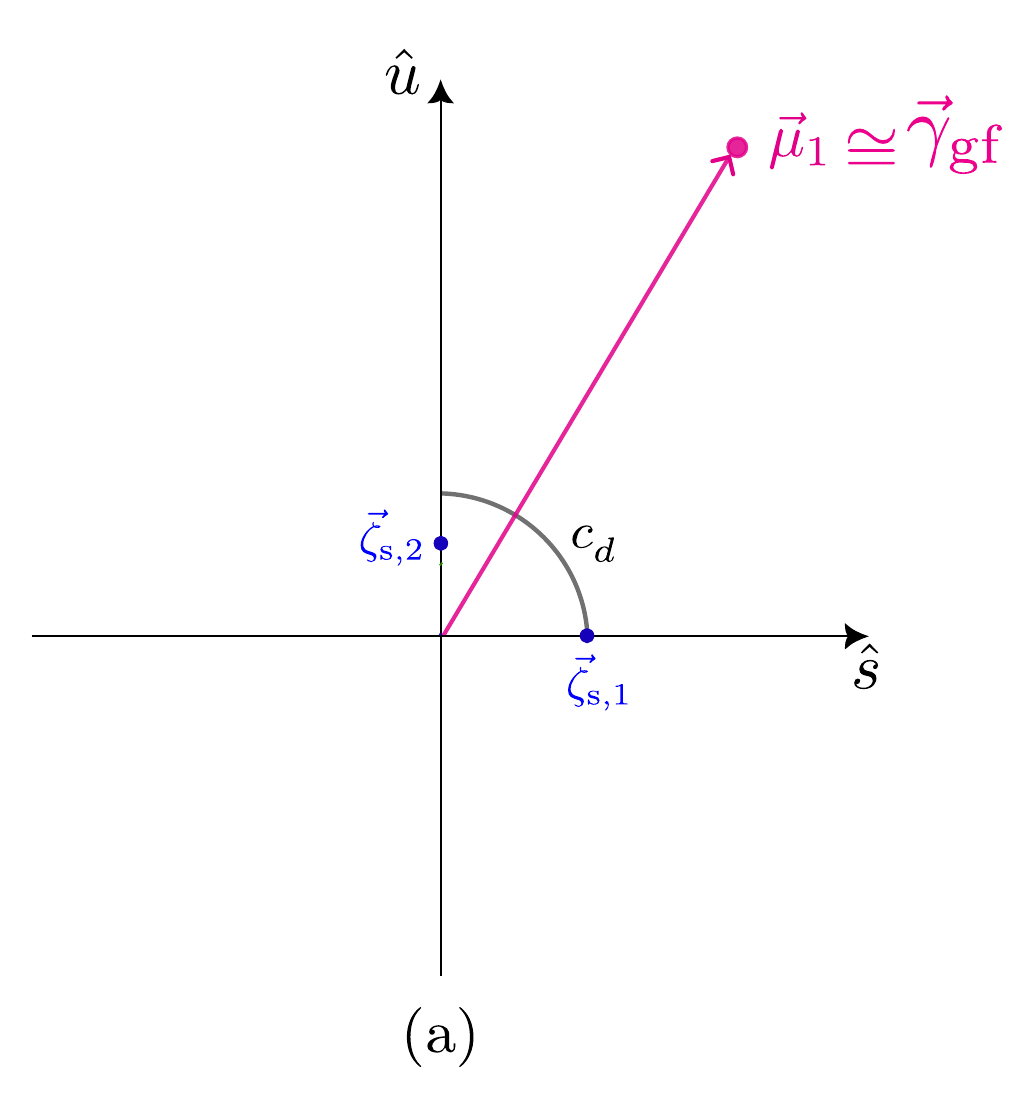}
    \hspace{0.01\linewidth}
    \includegraphics[width=0.48\linewidth]{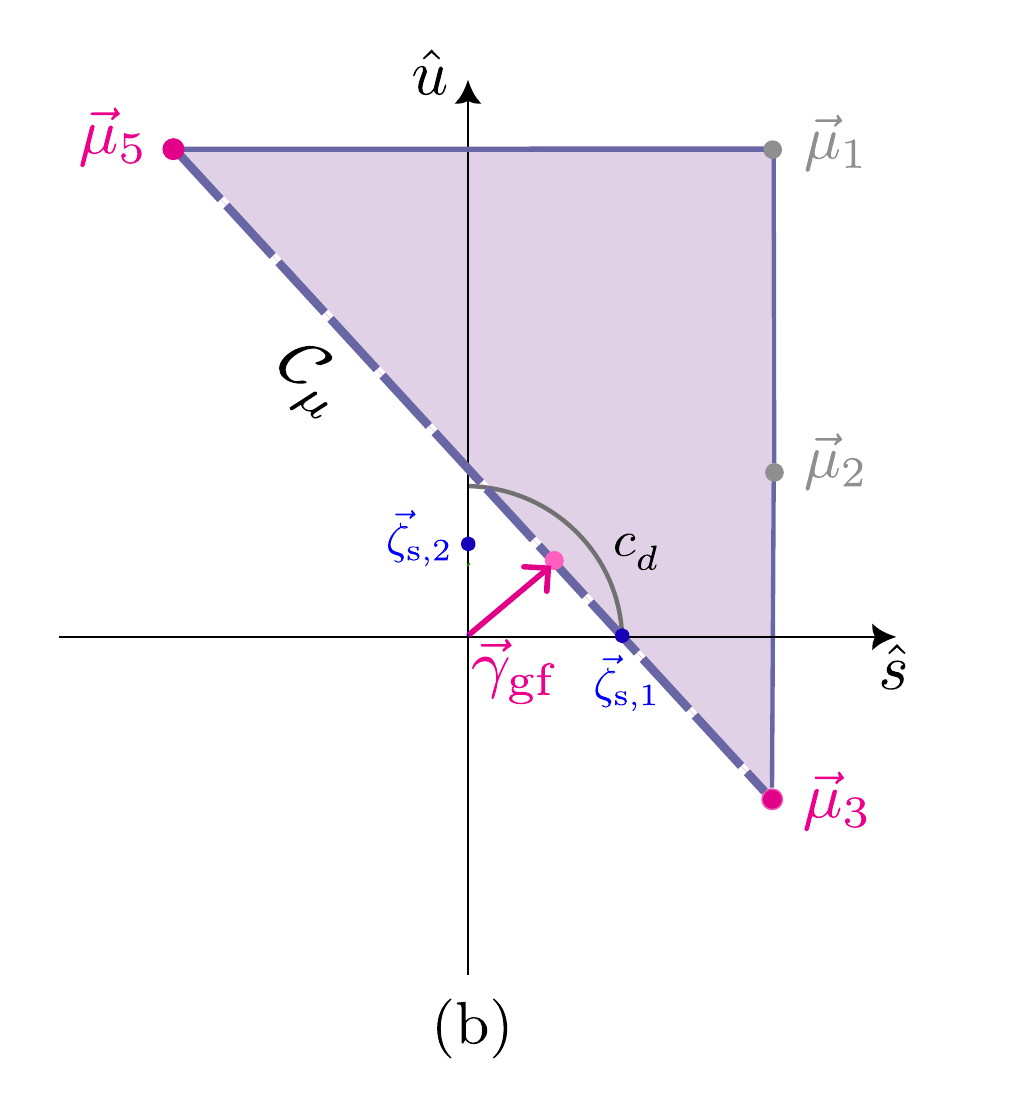}
    
    \caption{Examples of gradient flows in the tangent space, in setups where the elementary EFT strings are $\vec{\zeta_{\s,1,2}}$ (in blue). (a) When there is only one dominant potential term, $\vec{\mu}_1$, it coincides with the gradient flow trajectory, i.e.  $\vec{\gamma}_{\text{gf}} = \vec{\mu}_1$. (b) When the convex hull $\mathcal{C}_{\mu}$ is made of two competing dominant terms, $\vec{\mu}_3$ and $\vec{\mu}_5$, the gradient flow trajectory $\vec{\gamma}_{\text{gf}}$ coincides with the minimal distance direction to the origin. Notice that it is in the first quadrant even if the two competing terms are not.
    \label{fig:Vdom}}
\end{figure}

The two different cases are illustrated in \figref{fig:Vdom}. The loophole pointed out in \cite{Calderon-Infante:2022nxb} occurs in this second case involving competing terms of the potential and goes as follows. As shown in Figure \ref{fig:Vdom}, even if the individual terms of the potential satisfy the bound \eqref{eq:dsconj}, the convex hull of these terms can cut the ball of radius $c_d$ and therefore yield a gradient flow trajectory with $\gamma<c_d$ that accelerates asymptotically.

\subsubsection*{Validity of the EFT}
 
In the previous section, we saw that for a string to be fundamental, i.e. not resolvable within the regime of validity of the EFT, its tension must be larger than the cut-off scale of the theory \eqref{eq:fundamental}. In the presence of accelerated expansion, this in turn implies that the scale of the string tension needs to be heavier than the Hubble scale, since otherwise it would imply the breakdown of the EFT. In our setup \eqref{eqn:eftaction} in the slow-roll approximation\footnote{Despite being a relation that is usually attributed to the slow-roll approximation, the same scaling holds asymptotically for the cosmological solutions obtained when a scalar field is rolling down an exponential potential \cite{Rudelius:2022gbz,Shiu:2023fhb}, as happens asymptotically for the scaling solutions we are working with.}, this reads
\begin{equation} \label{eq:hubblevsstringscale}
    H(\varphi) = \sqrt{\frac{V(\varphi)}{3}} \lesssim T_{\s}^{1/2}\, .
\end{equation}
If strings become too light, the EFT would need to include all the stringy modes, which is not possible in a local quantum field theory description, so the EFT would break down.

Using the scaling vectors, the problem of the relative scales of strings and Hubble rate in \eqref{eq:hubblevsstringscale} just becomes a matter of comparing along the gradient flow the corresponding vectors $\vec{\zeta}_{\s,i}$ and $\vec{\zeta}_{H,\text{gf}}= \frac{1}{2}\vec{\gamma}_{\text{gf}}$, namely in order for the EFT strings to be heavy enough we should ask that for each of them 
\begin{equation} \label{eq:rhos}
\rho_{\s,i}\equiv \frac{\vec{\zeta}_{\text{s},i} \cdot \vec{\zeta}_{\text{H},\text{gf}}}{|\vec{\zeta}_{\text{H},\text{gf}}|^2} <1 \, .
\end{equation}
Due to the \n structure, with a field metric that is diagonal in the case of a leading monomial in the Kähler potential (hence restricting to a growth sector where this is the case), an elementary EFT string \eqref{eq:elementary-string} satisfies 
\begin{equation}
   \left(\zeta_{\s,i}\right)^{i} = \frac{1}{\sqrt{2n_i}} \, ,
\end{equation}
and the other components $\left(\zeta_{\s,i}\right)^{j\ne i}$ vanish, as in \eqref{eq:zetastr}. 
At the same time, the EFT string tension is bounded from below by the species scale $\Lambda_\text{sp}$ \cite{Martucci:2024trp,Grieco:2025bjy}. The latter also acts as an UV cut-off, with the IR cut-off given by the Hubble scale instead. Hence there is a hierarchy of scales given by
 \begin{equation}\label{eq:hierarchy}
     H\lesssim \Lambda_\text{sp}\lesssim T_\text{s}^{1/2}.
 \end{equation}
To keep track of this, it is also useful to define a $\rho_\text{sp}$ for the species scale and impose

\begin{equation}\label{eq:rho-sp}
\rho_{\text{sp}}\equiv\frac{\vec{Z}_\text{sp} \cdot \vec{\zeta}_{\text{H},\text{gf}}}{|\vec{\zeta}_{\text{H},\text{gf}}|^2} <1.
\end{equation}

\subsubsection*{Relation between particle towers and potential}

In this section, we want to show that in 4d \n theories, the scaling of the potential with the mass of the leading tower is bounded by the discrete details of the theory $n_i$ and $d_i$, differentiating the situations in which there is only one leading term in the potential, or two competing ones. The details of the computations can be found in Appendix \ref{app:B}, and here we just state the resulting bounds. 

As we explained above, in 4d \n theories both the $\zeta$-vectors of particles and the potential -- respectively, $\vec{\zeta}_{\t}$ and $\vec{\zeta}_V =2\vec{\zeta}_{\b}$ -- lie on the same lattice, generated by the $\zeta$-vectors of the elementary EFT strings $\zeta_{\s}$
\begin{equation}
    \vec{\zeta}_{V} = \sum_{i=1} w_{i,V} \vec{\zeta}_{\s_i} \, , \quad \vec{\zeta}_{\t} = \sum_{i=1} w_{i,\t} \vec{\zeta}_{\s_i} \, , 
\end{equation}
where $w_{i,V} = - 2\tilde{r}_i$ only depends on EFT data, and $w_{i,\t} =0, 1,2,3$ \cite{Lanza:2020qmt}. 

From \eqref{eq:lambdaV} and using \eqref{eq:zetastr}, it is then possible to define the relative scaling of the potential with respect to the mass of the tower, $\lambda$, purely in terms of EFT data and the scaling weights $w_{i,\t} = 0, 1, 2,3$.  In the presence of a single leading potential term for which $\vec{\gamma}_{\text{gf}} = \vec{\zeta}_{V}$, this can be done  as
\begin{equation} \label{eq:lambdaVt}
    \lambda = \frac{|\vec{\zeta}_{V}|^2}{\vec{\zeta}_V \cdot \vec{\zeta}_{\t}} =  - \frac{2 \sum \tilde{r}_{i}^2/n_i}{ \sum \tilde{r}_i  w_{i,\t}/n_i} \, ,
\end{equation} 
where for the gradient flow to be admissible $w_{i,V}\geq0$ with at least one that is non-vanishing, hence $\tilde{r}_i\leq0$ with at least one strictly negative.

Bounds on $\lambda$ can be obtained by assuming the lattice structure of the $\zeta-$vectors and the Integral Scaling Relation \eqref{eq:integral-scaling}. In the case of a single leading term these imply that 
\begin{equation} \label{eq:lambda1}
    \frac{2}{3} \leq \lambda \leq 2 \sum_i n_i \, .
\end{equation}
Notice that the lower bound is precisely the one found around \eqref{eq:lambda-species-bound} for $n=1$ decompactified dimension, while the upper bound in general cannot be further constrained if we do not make further assumptions on the sum of $\{n_i\}$, i.e. the degree $d_K$ of the Kähler potential.  

In the presence of two competing terms in the potential, the $\zeta-$vector associated to the asymptotic trajectory can also be expanded in the basis of elementary strings $\{ \vec{\zeta}_{\s_i} \}$, this time in terms of rational coefficients $k_i \in \mathbb{Q}$ instead of integer ones. The explicit expression of the $k_i$ in terms of the discrete data of the theory is given in \eqref{eq:ki}, but the important point is that in this basis, the $\lambda$ can be still expanded analogously to \eqref{eq:lambdaVt} with $\tilde{r}_i \to k_i$. The bounds on $\lambda$ for two leading terms in the potential assuming the lattice structure and the Integral Scaling Relation are
\begin{equation} \label{eq:lambda2}
    \frac{1}{3 \sqrt{\max_{i<j}(n_in_j)\sum \frac{\mathcal{R}_i^2}{n_i} \sum \frac{1}{n_i}}} \leq \lambda \leq 2 \sum_i n_i \, , 
\end{equation}
where $\mathcal{R}_i = \max(n_i+2d_{i-1},2d_i-n_i)$ is such that $\mathcal{R}_i \leq|\tilde{r}_i| \leq \mathcal{R}_i$. So, the bounds only depend on the values of $d_i$ and $n_i$. Notice, moreover, that the lower bound for two competing terms in the potential is generally lower than the one for a single term only, while the upper bounds are the same. 

We remind that these bounds for $\lambda$ only rely on the lattice structure of the $\zeta-$vectors in a 4d \n theory, and on the Integral Scaling Relation for the particles, so they are in general looser than \eqref{eq:lambdabound}. They can become stronger if we put additional constraints. For example, if we assume the sharpened bound $\frac{1}{\sqrt{2}}\leq \alpha \leq \sqrt{\frac{3}{2}}$ \cite{Etheredge:2022opl} and the fact that $|\vec{\zeta}_V| \geq \sqrt{6}$ in order for the single-term potential to be positive, then we can get 
\begin{equation}
    2 \leq \lambda \leq \frac{2 \sum n_i}{\sqrt{n_{\text{min}}}} \, .
\end{equation}
Of course, in cases where the no-scale structure is at work, the potential can be positive even if $|\vec{\zeta}_V|<\sqrt{6}$, so the lower bound above does not apply. Notice that it also does not apply for two leading terms, as the potential can be positive even for $|\vec{\gamma}_{\text{gf}}|\leq \sqrt{6}$. On the other hand, the upper bound would also apply in these other situations, as it does not rely on any numerical value for $|\vec{\zeta}_V|$. 

\subsection{\n EFT strings and potential: a bottom-up analysis}
Using the machinery outlined above, we now want to check from a bottom-up perspective\footnote{ The only UV input in what follows is assuming the existence of  physical states populating each EFT string charge.} whether the two conditions in Section \ref{sec:eftvalidity} can be satisfied simultaneously, namely whether the potential can yield asymptotic expansion ($\gamma < \sqrt{2}$) while keeping the EFT strings heavier than Hubble ($\rho_{\s}<1$).
In order to do so, we will restrict to a 2-moduli setting for simplicity and only assume bottom-up properties associated with \n\, theories -- namely, the presence of EFT strings $\vec{\zeta}_{\s,1}, \, \vec{\zeta}_{\s,2} $ and membranes $\{ \vec{\zeta}_{\b,i} \}$\footnote{For simplicity of notation, in this section we omit the subscript `b' from the component $w^{i}_{\b}$ of the brane scaling vector.} lying on the same lattice,
\begin{equation} \label{eq:vecb}
    \vec{\zeta}_{\b,i} = w_1^{\, i} \, \vec{\zeta}_{\s,1} + w_2^{\, i} \,  \vec{\zeta}_{\s,2}  \to \begin{pmatrix}
        w_1^{\, i} \\ w_2^{\, i}
    \end{pmatrix}\, , 
\end{equation}
together with the fact that scalar potential and elementary membrane tensions are connected by the relation $2V = Q_{\b}^2 \sim T_{\b}^2$ \cite{Lanza:2020qmt}, hence
\begin{equation}
    \vec{\zeta}_{H,\text{gf}} = \left( \vec{\zeta}_{\b,\text{dom}} \cdot \hat{t}_{\text{gf}} \right) \hat{t}_{\text{gf}} \, .
\end{equation}

In order to show that there are examples of theories with accelerated expansion without problematic light strings, we will start from considering the case (I) with only a single leading term of the potential, and then show some example with two competing terms (II), which are the only ones that may give accelerated expansion in explicit string-theory examples.

\subsubsection*{(I): $V_{\text{leading}}=V_{\bold{l}_1}$}

\begin{figure}[t]
    \centering
    \includegraphics[scale=0.4]{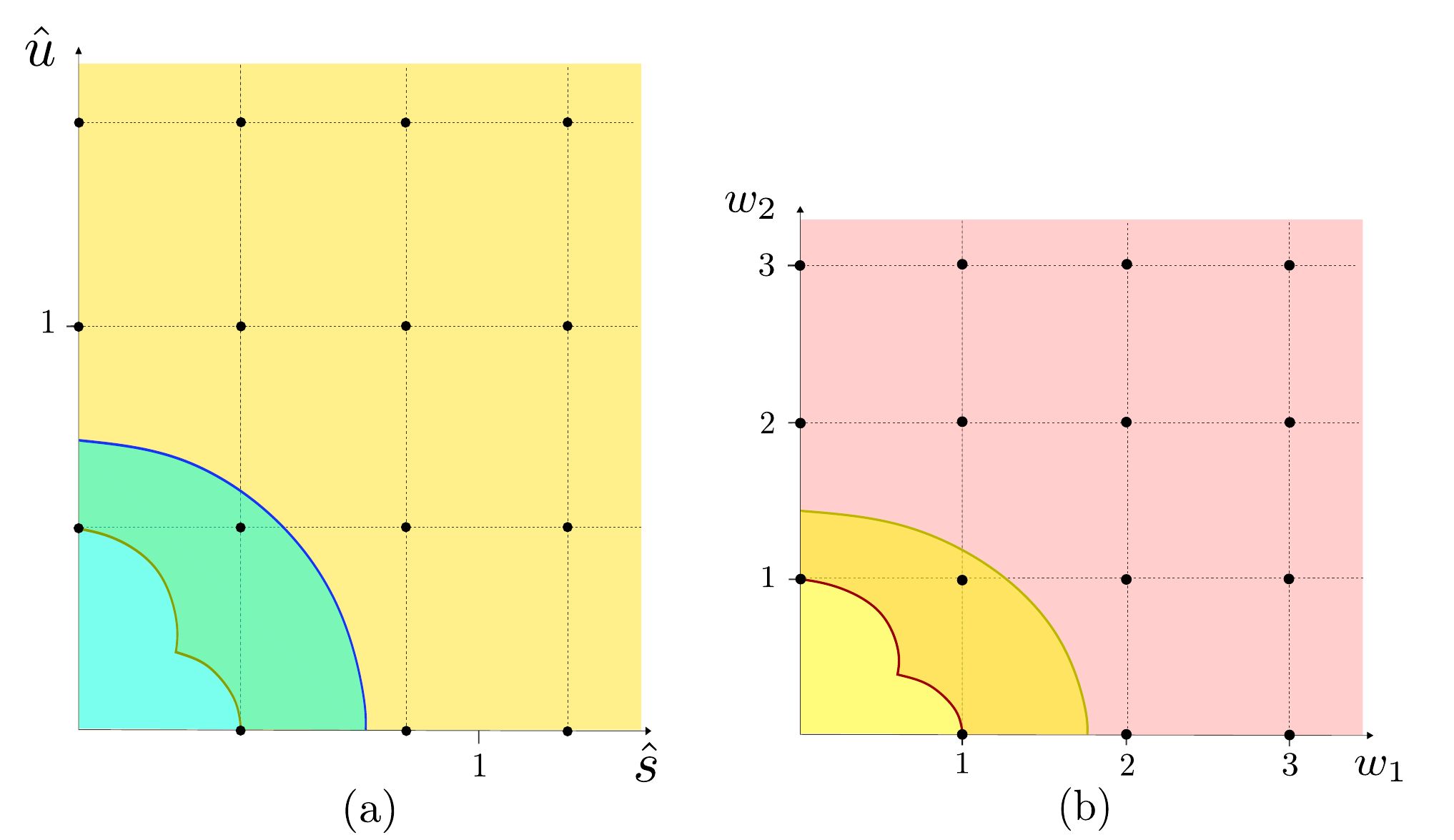} 
    
    \caption{Example for $n_1=3$ and $n_2=2$. (a) Lattice points, in the dual saxion space, for the Hubble scale $\zeta-$vectors belonging to the region for which the strings are heavier than the Hubble scale (yellow), and  the region for which $\gamma < \sqrt{2}$ (blue). In green, their intersection giving viable lattice points for controlled accelerated expansion. (b) same as the previous picture but in the basis of scaling weights $\omega$, giving a square lattice. The region for which $\gamma < \sqrt{2}$ here is in yellow, while the region for which the strings are heavier than the Hubble scale is in red. Their intersection is in orange.}
    \label{fig:ellipses}
\end{figure}

When a single term of the potential is dominating asymptotically, in order to give an admissible solution it must belong to $\mathcal{E}_{\text{light}}$, i.e. the corresponding $\zeta$-vector must lie in the first quadrant, $w_1,\,w_2 \geq 0$ with (0,0) excluded \cite{Calderon-Infante:2022nxb}. Imposing \eqref{eq:alphas} for the two elementary strings amounts to selecting the region 
\begin{equation}
    \begin{cases}
    \frac{w_1(1-w_1)}{2n_1} < \frac{(w_2)^2}{2n_2} \, , \\
    \frac{w_2(1-w_2)}{2n_2} < \frac{(w_1)^2}{2n_1} \, , \\
    w_1>0 \, , \\
    w_2 >0 \, , 
\end{cases}
\end{equation}
which we show in Figure \ref{fig:ellipses} for $n_1=3$ and $n_2=2$,
in two kinds of representations. In (a), the points lie in the lattice of elementary EFT strings, which in general is a rectangular one. This is the usual representation that appears in the literature (see e.g. \cite{Calderon-Infante:2022nxb,Grieco:2025bjy,Grieco:2026yip}). In (b), on the other hand, we plot the same in the square lattice of the $w$'s, in which each cell has area 1. We introduce this representation because it will be more suited for what we will say in the following.

Since for each lattice point $\{ w_1,w_2 \}$ the left hand side of the first two inequalities is non-positive, while the second is non-negative (remember that $n_1, \, n_2>0$), we can conclude that independently on the degrees of the Kähler potential and the moduli-space trajectory selected by the scalar potential, the strings are always heavier than the Hubble scale\footnote{This conclusion does not depend on the number $N$ of moduli, as in general the system is made of $N$ inequalities, each one looking like
\begin{equation}
    \frac{w_i (1-w_i)}{2n_i} < \sum_{j \ne i, j = 1}^{N} \frac{(w_j)^2}{2n_j} \, ,
\end{equation}
for which what said above also applies.}.

On the other hand, the requirement of accelerated expansion $\gamma<\sqrt{2}$, which in terms of the $w_i$ reads
\begin{equation} \label{eq:accexp}
    \frac{(w_1)^2}{2n_1} + \frac{(w_2)^2}{2n_2} < \frac{1}{2} \, ,
\end{equation}
allows few lattice points for the Hubble scale to belong to the region associated with $\gamma<\sqrt{2}$, and only for some specific choices of $\{n_1,n_2\}$, that is, $\{ \geq 2 , \geq 2 \}$ with the case $\{ 2,2 \}$ excluded - i.e. if the degree of the Kähler potential $d_K=\sum_i n_i$  is $d_K > 4$. 

It is then clear that the set of allowed choices of potentials (i.e. brane content) which yield a consistent EFT compatible with both accelerated expansion and heavy strings is just the intersection of the previous two regions, and it is not empty for the same set of $n$'s that give $\gamma < \sqrt{2}$ -- that is, all the choices allowing for accelerated expansion give a consistent EFT in terms of the string scale. 

 On the other hand, none of these states actually belong to the spectrum derived from top-down examples, as the individual potential terms are bounded. First of all, as noted around \eqref{eq:accexp}, the two-moduli limits of F-theory compactifications on $CY_4$ (with a single term dominating the potential) cannot give rise to accelerated expansion as the degree of the Kähler potential is $d_K=4$ and is incompatible with the set of $\{ n_1, n_2\}$ giving accelerated expansion. 

Even assuming in the two-moduli example the degree of the Kähler potential to be $d_K=7$ as in M-theory on $G_2$ manifolds, the only allowed Hubble vector giving accelerated expansion would be $(w_1,w_2) =(1,1)$, which does not obviously correspond to any of the known allowed fluxes in this top-down setup, although $G_2$ manifolds with fluxes are admittedly less explored. The case with more than two moduli constrains even more the choice of allowed $w$'s, and is then again excluded.

In the one-modulus case there are less constraints, and a priori $w_1=1$ is always allowed for $n_1 \geq 2$.  

Let us notice that whether or not the sign of the potential is actually positive given these $\zeta-$vectors depends on whether the two moduli $\{\varphi^i \}_{i=1,2}$ we consider as contributing to the potential are all the ones present in the theory or just a subset. Indeed, in the case where there are additional moduli that enforce the no-scale structure, the potential will by definition be positive, as 
\begin{equation}
    V = e^{K} K^{i \bar{j}} D_i W D_{\bar{j}} \overline{W} \, ,
\end{equation}
which is what happens in the F-theory examples that we consider in Section \ref{sec:Ftopdown}.

On the other hand, if these are the only moduli of the theory, then 
\begin{equation}
    V = \frac{1}{2} \left( |\nabla T|^2 - \frac{3}{2}T^2\right) = \frac{1}{4} \left( \frac{|\zeta_V|^2}{2}  -3 \right) T^2 > 0 \iff |\zeta_V|^2 > \sqrt{6} \, .
\end{equation}
Hence a potential with a single leading term with $|\vec{\zeta}|_V\leq \sqrt{6}$ would have a negative sign, such that asymptotic accelerated expansion would be prevented as remarked in \cite{Rudelius:2021azq,Hellerman:2001yi}.
Other no-go's for acceleration were also given in \cite{Hebecker:2023qke}, assuming specific stringy energy sources. This is not incompatible with our results here, since in this section we are taking a bottom-up approach. 

\subsubsection*{(II): $V_{\text{leading}}=V_{\bold{l}_1} + V_{\bold{l}_2}$}

In this setup, the most general question we may ask is: given two brane vectors lying in the EFT strings lattice, like 
\begin{align}
     \vec{\zeta}_{\b,1} &= \text{w}_{1} \, \vec{\zeta}_{\s,1} + \text{w}_{2} \, \vec{\zeta}_{\s,2} \, , \\
     \vec{\zeta}_{\b,2} &= \omega_{1} \, \vec{\zeta}_{\s,1} + \omega_{2} \, \vec{\zeta}_{\s,2} \, , 
\end{align}
what are the possible (non-trivial) pairs $\{ \vec{\text{w}}, \, \vec{\omega}\}$ that give accelerated expansion? 

First of all, we need to ensure that the given pair yields an admissible gradient flow trajectory. Given that the tangent vector to the gradient flow trajectory $\hat{t}_{\text{gf}}$ is parallel to the Hubble rate $\zeta-$vector
\begin{equation} \label{eq:gammaHgf}
    \vec{\zeta}_{\text{H},\text{gf}} =  \frac{\text{w}_2 \omega_1 - \text{w}_1 \omega_2}{\sqrt{2} \left(n_2 (\text{w}_1-\omega_1)^2 + n_1 (\text{w}_2-\omega_2)^2\right)} \left( \sqrt{n_1} (\text{w}_2-\omega_2) , - \sqrt{n_2} (\text{w}_1-\omega_1) \right) \, ,
\end{equation}
this implies that $\vec{\zeta}_{\text{H},\text{gf}}$ must lie in the first quadrant, i.e. $\zeta_{\text{H},\text{gf}}^{\, \,1,2}>0$.  

On top of this consistency condition, we need to ask, as above, that 
\begin{equation}
    |\vec{\zeta}_{\text{H},\text{gf}}|<\frac{1}{\sqrt{2}}, \quad \alpha_{\s,1,2}<1 \, .
\end{equation}
This system of inequalities can be solved, but given the high number of variables the results are difficult to present in full generality. Since the main objective is to show that the pattern is richer than in the case (I) -- namely, that the absence of light strings along gradient flow trajectories is not guaranteed -- we just focus on one example. 

We choose the values $n_1=4$,  $n_2=2$ which are outside of the CY regime. Assuming moreover that $d_i=n_i'$, the allowed brane vectors $\vec{\zeta}_{\b} = (w_{\b,1}, w_{\b,2})$ are restricted to have components
\begin{equation} \label{eq:wbounds}
    -4 \leq w_{\b,1} \leq 4 \, , \quad -10 \leq w_{\b,2} \leq 10\, , \quad -6 \leq w_{\b,1} + w_{\b,2} \leq 6 \, .
\end{equation}
We then fix the first vector to have $\text{w}_1=3$ and $\text{w}_2=-5$, such that $|\vec{\zeta}_{\b,1}|=\sqrt{\frac{59}{8}} > \sqrt{\frac{3}{2}}$, enforcing that the corresponding term in the potential is positive. For the same reason, we also restrict to $|\vec{\zeta}_{\b,2}|>\sqrt{\frac{3}{2}}$.

\begin{figure}[t]
    \centering
    \includegraphics[width=0.48\linewidth]{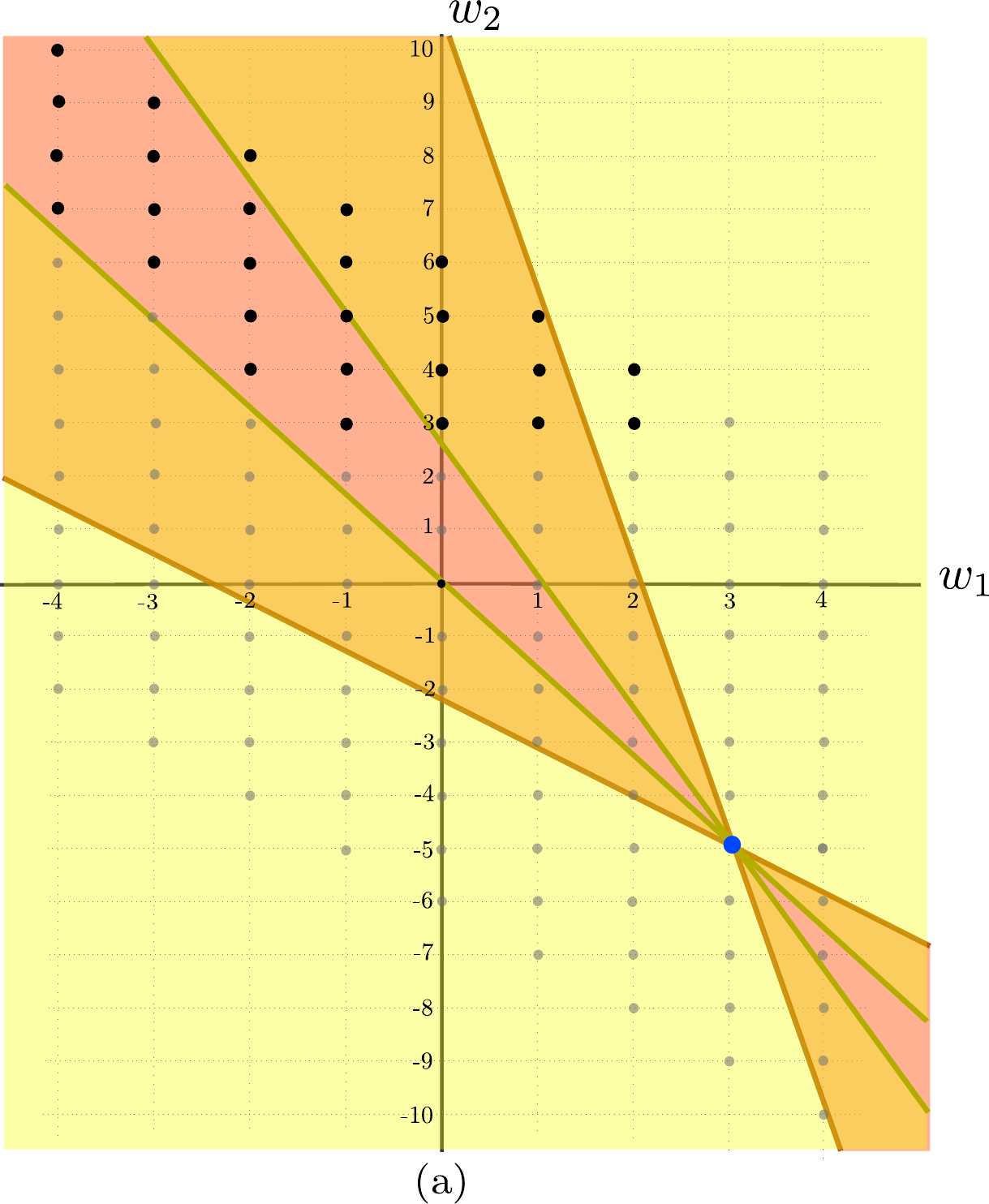}
    \hspace{0.01\linewidth}\includegraphics[width=0.48\linewidth]{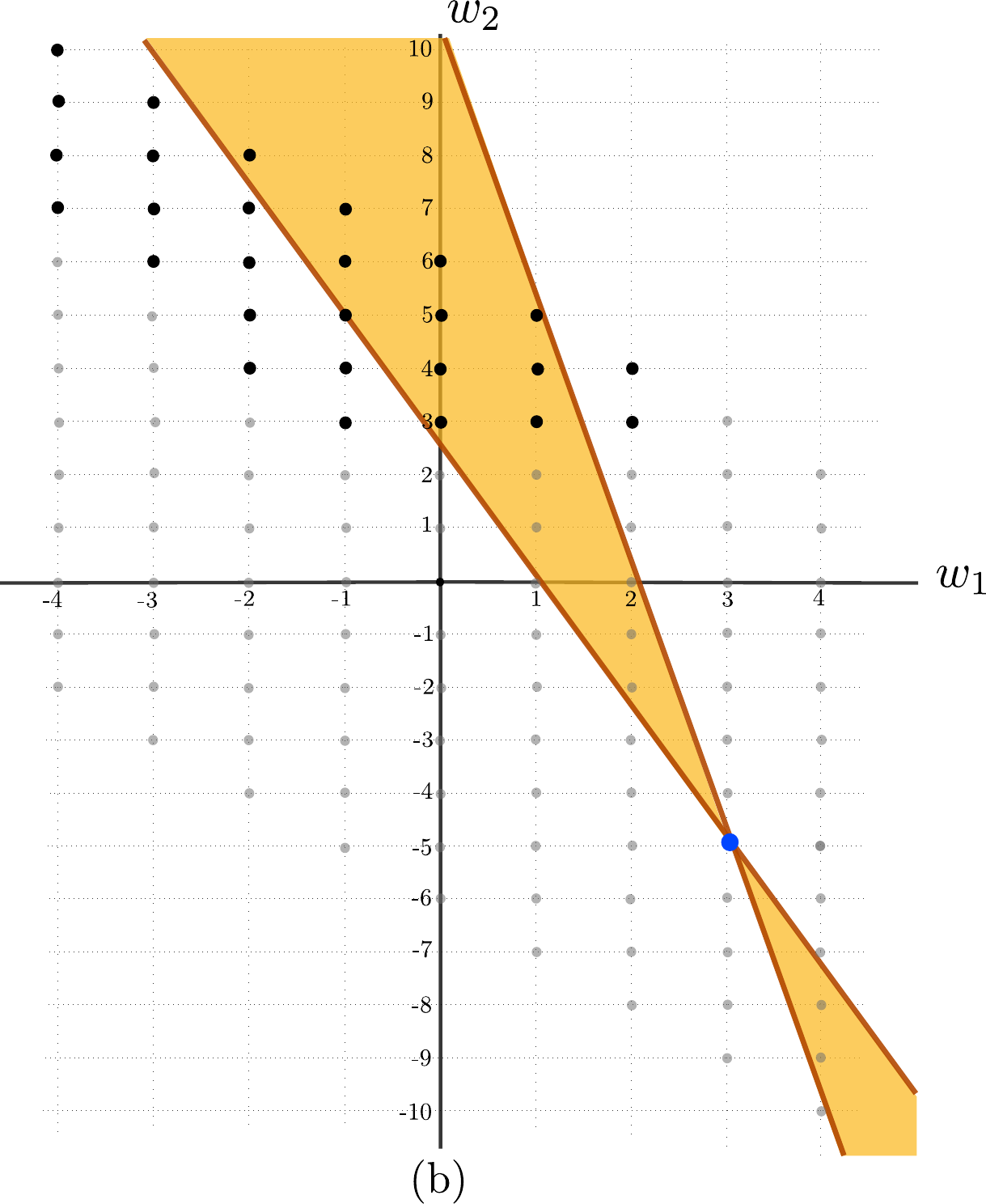}
  
  \caption{(a) In yellow (with green boundary), vectors $\vec{\zeta}_{\b,2}$ - bold black points - that are leading together with $\vec{\zeta}_{\b,1}=(3,-5)$ (blue point) for $(n_1,n_2)=(4,2)$ and give a gradient flow trajectory along which the string tensions decay slower than the Hubble scale. In the red region (with orange boundary), vectors $\vec{\zeta}_{\b,2}$ that together with  $\vec{\zeta}_{\b,1}=(3,-5)$ for $(n_1,n_2)=(4,2)$ give a gradient flow trajectory staying in a region of vanishing potential and may lead to accelerated expansion. The portion of their intersection populated by allowed potential terms gives (b), in orange, containing vectors $\vec{\zeta}_{\b,2}$ that together with $\vec{\zeta}_{\b,1}=(3,-5)$ for $(n_1,n_2)=(4,2)$ give accelerated expansion without light strings. The gray points in both (a) and (b) are lattice points that, together with $\vec{\zeta}_{\b,1}=(3,-5)$, do not correspond to a total gradient flow in the first quadrant, hence are discarded.} 
    \label{fig:region1}
\end{figure}

In this setup, we want to show what happens to the relevant scales of the theory as we change the position of $\vec{\zeta}_{\b,2}$ in the lattice within the aforementioned ranges \eqref{eq:wbounds}. Such allowed vectors $\vec{\zeta}_{\text{b},2}$ are shown as black dots in Figure \ref{fig:region1}(a,b).

First of all, the requirement of an admissible gradient flow trajectory is given by a
system of inequalities in $\{ \omega_1, \omega_2 \}$ which, for the example at hand, can be reduced to
\begin{equation} \label{eq:reg1}
(\omega_1 - \text{w}_1) \left(\omega_2 - \frac{\text{w}_2}{\text{w}_1} \, \omega_1 \right) < 0  \, .
\end{equation}

Then, in order to have accelerated expansion, the terms that can be leading at the same time as $\vec{\zeta}_{\b,1}$ must satisfy -- for our choice of parameters
\begin{equation} \label{eq:reg2}
\begin{cases}
    n_1(\text{w}_2-\omega_2)+ \text{w}_1  (\text{w}_1 \omega_2 - \text{w}_2 \omega_1) + \sqrt{n_2 \text{w}_1^2 + n_1(\text{w}_2^2 -n_2)}|\text{w}_1 -\omega_1|> 0 \, , \\
    \omega_2 (n_1 - \text{w}_1^2) - n_1 \text{w}_2+\text{w}_1\text{w}_2\omega_1 + \sqrt{n_2 \text{w}_1^2 + n_1(\text{w}_2^2 -n_2)}|\text{w}_1 -\omega_1| > 0 \, ,
\end{cases}
\end{equation}
which selects the red region in  Figure \ref{fig:region1}(a).  

Finally, we need to impose that the Hubble scale vanishes faster than the string tension. For our choice of parameters, this amounts to $\vec{\zeta}_{\b,2} \ne \vec{\zeta}_{\b,1}$ and
\begin{equation} \label{eq:reg3}
        (\omega_1 -\text{w}_1) \left( \omega_2 - \frac{\text{w}_2}{\text{w}_1} \, \omega_1 \right) \geq 0 \text{ or }  (\omega_1 -\text{w}_1) \left(\omega_2  \frac{\text{w}_2}{\text{w}_1-1} \, (\omega_1-1) \right) \leq 0 \, .
\end{equation}
In particular, we see that, unlike the case (I), not all the choices of leading pairs in the potential lead to the two elementary EFT strings being light, the only viable $\vec{\zeta}_{\b,2}$ being the ones in the yellow region in Figure \ref{fig:region1}(a).

Superimposing the three regions \eqref{eq:reg1}, \eqref{eq:reg2} and \eqref{eq:reg3} one gets that the allowed $\vec{\zeta}_{\b,2}$ are the ones in the orange region of Figure \ref{fig:region1}(b), and we see that they do not coincide with the ones that give accelerated expansion (in the red region of Figure \ref{fig:region1}(a)).

Notice that in order for the analysis above to work, one must ensure that along the gradient flow trajectory the potential indeed vanishes asymptotically and stays positive. This always happens if both the leading potential terms are associated to $\vec{\zeta}_{\b} > \sqrt{\frac{3}{2}}$ as above. When one of the two does not does not satisfy this inequality, and hence one of the potentials term becomes negative, the behaviour of the potential along the direction \eqref{eq:gammaHgf} needs to be explicitly checked. Indeed, in this situation the gradient flow direction does not need to coincide with $\hat{t}_{\text{gf}}$, and moreover the potential may diverge negatively along the correct asymptotic trajectory. In the case at hand, anyway, the positivity of the potential is guaranteed for all the viable terms. 

Hence, we see that it is possible a priori to satisfy both conditions at the same time, and hence to have an EFT in 4 dimensions with \n\,describing consistently accelerated expansion. Hence, if there is any obstruction for asymptotic acceleration, it must clearly contain some further UV input. 

In particular, let us mention that the possibility of having accelerated expansion in a well-controlled EFT is still consistent with the claims of \cite{Etheredge:2026cha}. Indeed, their result on the impossibility of violating the strong dS bound relies on specific properties of the scaling vectors of the membranes, which are more restrictive than the general bottom-up ones we consider. 

\section{Top-down results from F-theory and Type II string theories} \label{sec:topdown}

The goal of this section is to focus on concrete \n\, Type II and F-theory flux compactification in the presence of a positive runaway potential, in order to study explicit field trajectories that arise in the string landscape. We are particularly interested in the specific relations between particles, strings and potential/Hubble scale. We add, to the top-down analysis of particles and EFT strings in \cite{Grieco:2025bjy}, the behaviour of brane tensions and potential in a systematic way in different asymptotic limits, in particular to investigate whether they can yield asymptotic acceleration satisfying the conditions of  Section \ref{sec:lightstrings}. We will focus on the explicit 2-moduli limits studied in \cite{Grimm:2019ixq,Calderon-Infante:2022nxb} and find that the potential candidates yielding acceleration actually have too light strings, closing the loophole of \cite{Calderon-Infante:2022nxb}.  

\subsection{EFT strings and membranes in 2-moduli asymptotic F-theory limits} \label{sec:Ftopdown}

\subsubsection*{Type IIB at large complex structure and weak string coupling} \label{sec:IIB1}

The first type of limits we can consider, following \cite{Calderon-Infante:2022nxb}, are Type IIB CY$_3$ orientifolds at weak string coupling in the large complex structure regime, restricted to the moduli space of the dilaton and a single complex structure modulus. This corresponds to Sen weak coupling limit in the complex structure moduli space of F-theory on an elliptically fibered CY$_4$ \cite{Sen:1996vd}, followed by taking a single complex structure modulus to large value, i.e. a II$_{0,1} \to$ V$_{2,2}$ singularity enhancement for the CY$_4$ moduli space \cite{Grimm:2019ixq}.

We define respectively the axio-dilaton and said complex structure modulus to be the chiral fields $\varphi^1 = a_1 + is$ and $\varphi^2 = a_2 + iu$, and restrict to the growth sector $\mathcal{R}_{su} = \left\{ \frac{s}{u} > \delta, u > \delta, a_i < \beta \right\}$ in the F-theory framework, for arbitrary positive $\delta>1,\, \beta<1$.

The Kähler potential describing the physics on this two-dimensional slice of the moduli space is 
\begin{equation}\label{eq:IIBKahler}
    K(s,u) = - \log \left(s\right) - 3 \log \left(u \right) \, .
\end{equation}
From the Type IIB perspective, the elementary EFT strings in such a sector of the theory correspond to either D7 branes wrapping the whole CY threefold or to D3 branes wrapping a 2-cycle \cite{Lanza:2021udy}. Their tension in 4 dimensional Planck units is respectively
\begin{equation}\label{eq:IIBtensions}
    T_{\text{s},1}(s,u) = \frac{1}{2s} \, ,  \quad T_{\text{s},2}(s,u) = \frac{3}{2u} \, ,
\end{equation}
and the associated $\zeta$-vectors are
\begin{equation}
    \zeta_{\text{s},1} = \left( \frac{1}{\sqrt{2}}, 0\right)\, ,  \quad \zeta_{\text{s},2} = \left( 0,\frac{1}{\sqrt{6}} \right)\, .  
\end{equation}

A potential for these moduli can be sourced by turning on 3-form fluxes, either in the R-R or NS-NS sector. By expanding them in a basis of harmonic forms preserved by the orientifold, we define the corresponding components\footnote{We follow the  notation for fluxes as in \cite{Calderon-Infante:2022nxb}, in order to match the different terms in the scalar potential to the ones in the mirror dual Type IIA setup.} as follows
\begin{equation} \label{eq:fluxquanta}
    G_3 \to (f_0, \, f_2, \, f_4, \, f_6) \, , \quad H_3 \to (h_0, \, h_1, \, h_2, \, h_3) \, .
\end{equation}
The induced superpotential takes the form \cite{Grimm:2004uq} 
\begin{equation} \label{eq:wIIB}
    W(\varphi^1,\varphi^2) = f_6 - f_4 \varphi^2 + \frac{f_2}{2} \left(\varphi^2 \right)^2 - \frac{f_0}{6}  \left(\varphi^2 \right)^3 + \varphi^1 \left(h_0 - h_1 \varphi^2 + \frac{h_2}{2} \left(\varphi^2 \right)^2 - \frac{h_3}{6}  \left(\varphi^2\right)^3 \right) \, ,
\end{equation}
where we omit the contributions of non-geometric fluxes to the superpotential, which would depend on the overall volume. 

The corresponding EFT membranes, coming from D5-branes wrapping 3-cycles, have tensions \eqref{eq:tension} in which the dependence on axions and saxions is factorised. In this way, the $\zeta$-vectors are the same with or without axion stabilisation, and so can be read from the term of order zero in the axions, $T^{0}_\mathbf{q}$, which in each case is  
\begin{align}
    T_{\text{b},f_6}^{\text{IIB},0}(s,u) &= \frac{2 f_6}{\sqrt{su^3}} \, , &&\zeta^{\text{IIB}}_{\text{b},f_6} = \left( \frac{1}{\sqrt{2}},\sqrt{\frac{3}{2}} \right) \, , \label{eq:tension1}\\
    T_{\text{b},f_4}^{\text{IIB},0}(s,u) &= \frac{2f_4}{\sqrt{su}} \, , &&\zeta^{\text{IIB}}_{\text{b},f_4} = \left( \frac{1}{\sqrt{2}},\frac{1}{\sqrt{6}} \right) \, , \\
    T_{\text{b},f_2}^{\text{IIB},0}(s,u) &= f_2 \sqrt{\frac{u}{s}} \, , &&\zeta^{\text{IIB}}_{\text{b},f_2} = \left( \frac{1}{\sqrt{2}},-\frac{1}{\sqrt{6}}\right) \, , \\
    T_{\text{b},f_0}^{\text{IIB},0}(s,u) &= \frac{f_0}{3}\sqrt{\frac{u^3}{s}}  \, , &&\zeta_{\text{b},f_0}^{\text{IIB}} = \left( \frac{1}{\sqrt{2}},-\sqrt{\frac{3}{2}} \right) \, , \\
    T_{\text{b},h_0}^{\text{IIB},0}(s,u) &= 2h_0 \sqrt{\frac{s}{u^3}}\, , &&\zeta_{\text{b},h_0}^{\text{IIB}} = \left( -\frac{1}{\sqrt{2}}, \sqrt{\frac{3}{2}} \right) \, , \\
    T_{\text{b},h_1}^{\text{IIB},0}(s,u) &= 2 h_1 \sqrt{\frac{s}{u}} \, , &&\zeta_{\text{b},h_1}^{\text{IIB}} = \left( -\frac{1}{\sqrt{2}},\frac{1}{\sqrt{6}}\right) \, , \\
    T_{\text{b},h_2}^{\text{IIB},0}(s,u) &= h_2 \sqrt{su} \, , &&\zeta_{\text{b},h_2}^{\text{IIB}} = \left( -\frac{1}{\sqrt{2}},-\frac{1}{\sqrt{6}}\right) \, , \\
    T_{\text{b},h_3}^{\text{IIB},0}(s,u) &= \frac{h_3}{3}\sqrt{su^3}  \, , &&\zeta_{\text{b},h_3}^{\text{IIB}} = \left( -\frac{1}{\sqrt{2}}, -\sqrt{\frac{3}{2}} \right) \, . \label{eq:tension2}
\end{align}
The flux in the subscript of each tension denotes the leading term in each $\bold{\rho}_{\text{b},f_i}(\mathbf{q},a) = f_i + \cdots$ as in \eqref{eq:branes}, which is indeed the one contributing to the order-zero term in the axions. Axion stabilisation is important in order to determine, for each choice of fluxes, which are the $\rho_{\text{b}}$ functions that survive, so that we can obtain the corresponding set of non-vanishing $\zeta_{\b}$ vectors for the specific background at hand \cite{Calderon-Infante:2022nxb}. 

For a generic choice of fluxes and assuming that the Kähler modulus can be stabilised independently, the potential is    
\begin{equation} \label{eq:potential1}
V^{\text{IIB}}(\varphi^1, \varphi^2) = \frac{V_{f_6}}{s u^3} + \frac{V_{f_4}}{s u} + \frac{V_{f_2} u}{s} + \frac{V_{f_0} u^3}{s} + \frac{V_{h_0} s}{u^3} + \frac{V_{h_1}s}{u} + V_{h_2} su + V_{h_3} su^3 + V_{const} \, .
\end{equation}    
Using \eqref{eq:scalarpotentialn1}, each coefficient $V_i$ depends in general on the axions $a^1$ and $a^2$ and on the flux quanta \eqref{eq:fluxquanta}.\footnote{The constant contribution is such to enforce tadpole cancellation.} Solving first for the saxions trajectory and then studying the axion behaviour along that gradient flow, it can be shown that the leading terms determining the asymptotic trajectory are independent of the axions, which either get stabilised at finite values or remain as flat directions \cite{Calderon-Infante:2022nxb}.  More recently, a full dynamical analysis of the axion-saxion system was performed in \cite{Grimm:2025cpq}. Here they show that late-time solutions are fixed points of the dynamical system, generically with a constant axion-to-saxion ratio, except when the leading asymptotic potential vanishes at its minimum, in which case oscillatory axionic solutions can arise asymptotically. Subleading corrections that do not vanish at the minimum, such as the $\alpha'$ corrections considered in \cite{Grimm:2025cpq}, lift this degeneracy and render the oscillatory solutions unstable, driving the system towards a fixed point. So, it is consistent to just neglect the axion contributions.

Depending on which fluxes are turned on, we will get different asymptotic trajectories. The different choices of fluxes that can yield an asymptotic gradient flow trajectory along which the potential vanishes at infinite distance were studied in \cite{Calderon-Infante:2022nxb}.
We summarise in Table \ref{tab:IIBlargevollargecs} the different gradient flow trajectories $\vec{\gamma}_{\text{gf}}$ \eqref{eq:gammagf} for the allowed choices of fluxes as in \cite{Calderon-Infante:2022nxb}. For each one, we specify the exponential rate of the leading EFT string along the trajectory itself
\begin{equation} \label{eq:alphas}
    \alpha_{\s} = \frac{\vec{\zeta}_{\s} \cdot \vec{\gamma}_{\text{gf}}}{\vec{\gamma}_{\text{gf}}} \, ,
\end{equation}
and the relative exponential rate between the leading string and the Hubble rates along the selected trajectory, namely $\rho_{\s} = \frac{\alpha_{\s}}{|\vec{\zeta}_{H,\text{gf}}|} \equiv 2 \frac{\alpha_{\s}}{|\vec{\gamma}_{\text{gf}}|}$, which was already defined in \eqref{eq:rhos} and is smaller than one for a string that remains heavier than the Hubble scale in the asymptotic regime. As was pointed out already in \cite{Calderon-Infante:2022nxb}, in this setting there is a flux choice that is in tension with the strong dS conjecture and the TCC, with a value $|\vec{\gamma}_{\text{gs}}| = \sqrt{\frac{2}{7}}$. We can see in the table that this comes at the price of the leading string tension becoming lighter than the Hubble scale, as $\rho_{\s}=2>1$.

\begin{figure}[t]
    \centering
    \includegraphics[width=0.6\linewidth]{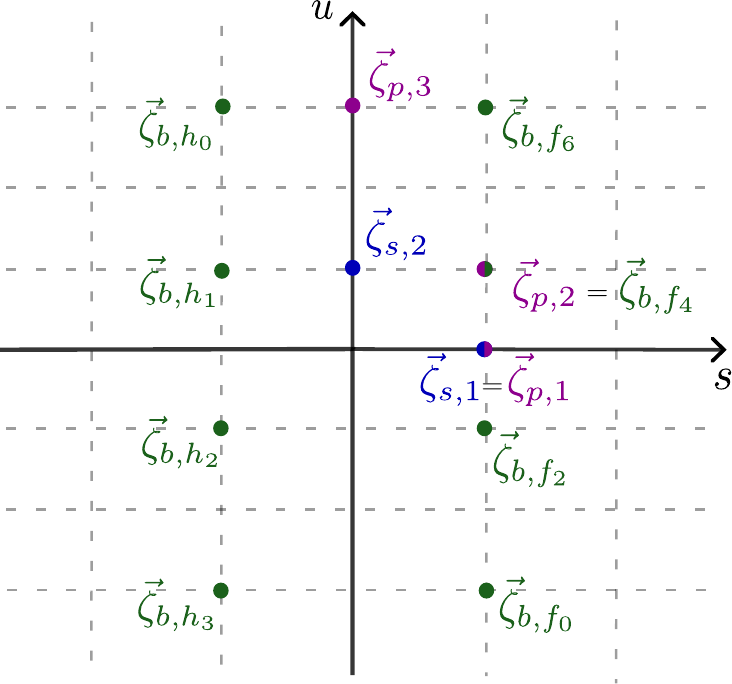}
    \caption{Particles (in purple), EFT strings (in blue) and EFT membranes (in green) for Type IIB CY$_3$ orientifold at weak coupling and large complex structure.}
    \label{fig:pbstypeIIB1}
\end{figure}

For these compactifications, the leading particle towers can be obtained from mirror symmetry with Type IIA orientifolds in the large volume limit, where the microscopic interpretation as KK and string towers is clearer. We defer the discussion of their microscopic interpretation to later in this section when discussing Type IIA examples. Particle towers have the following $\zeta$-vectors:
\begin{equation} \label{eq:towers}
    \zeta_{\text{p},1} = \left( \frac{1}{\sqrt{2}}, 0 \right)\, ,  \quad \zeta_{\text{p},2} = \left( \frac{1}{\sqrt{2}}, \frac{1}{\sqrt{6}}\right)\, ,  \quad \zeta_{\text{p},3} = \left( 0, \sqrt{\frac{3}{2}}\right)\, . 
\end{equation}
We show the lattice of the $\zeta$-vectors of particles, strings and membranes in Figure \ref{fig:pbstypeIIB1}. We also show in Table \ref{tab:IIBlargevollargecs} the behaviour of the leading terms of the potential with respect to the leading towers, quantified by the parameter $\lambda$ as $V \sim m_{\t}^{\lambda}$ as in \eqref{eq:lambdaV}. All the values of $\lambda$ in the table are consistent with the bounds \eqref{eq:lambda1} and \eqref{eq:lambda2}, which in this context read respectively $\frac{2}{3} \leq \lambda \leq 8$ for a single leading term and $\frac{1}{12} \leq \lambda \leq 8$ for two competing ones. Notice that in the pathological case, the value of $\lambda=\frac{2}{3}$ is inconsistent with the expectations from the Higuchi bound \cite{Higuchi:1986py}, i.e. $\lambda\geq 2$. 
We also add the value of the species scale \eqref{eq:spscale} in each gradient flow direction and its magnitude with respect to the Hubble scale $\rho_\text{sp}$, defined as \eqref{eq:rho-sp}. In the example leading to acceleration, the species scale becomes lighter than the Hubble scale, which is another signal of the pathological nature of the limit.

\begin{table}[htb!]
    \centering    
\begin{tabular}{ |c|c|c|c|c|c|c|c|c|c| } 
 \hline
 $V(s,u)\big|_{\text{leading}}$ & $\vec{\gamma}_{\text{gf}}$ &$|\vec{\gamma}_{\text{gf}}|$ & 
 $\rho_{\text{s}}$ & $\alpha_{\text{s}}$ 
 & $\Lambda_{\text{sp}}$  &$\rho_{\text{sp}}$& Leading tower &$\lambda$ 
 \\  \hline

$\frac{V_{f_4}}{s\,u}$ & $\left(\sqrt{2}, \sqrt{\frac{2}{3}}\right)$ & $2 \sqrt{\frac{2}{3}}$ & $ \frac{3}{4}$ & $\sqrt{\frac{3}{2}}$ 
& $M_{\text{Pl,10}}$ &  $\frac{3}{4}$ & $m_{\text{KK,X}}$ & $2$ 
\\ \hline

$\frac{V_{h_0} s}{u^3}$ & $\left( \frac{1}{\sqrt{2}}, \sqrt{\frac{3}{2}} \right)$  & $\sqrt{2}$ & $\frac{1}{2} $ & $\frac{1}{\sqrt{2}}$ 
& $M_{\text{Pl,11}}$ & $\frac{5}{6}$ & $m_{\text{KK,M-th}}$ & $\frac{4}{3}$ 
\\ \hline

$\frac{V_{f_6}}{s\, u^3} + \frac{V_{f_0}u^3}{s} $ & $\left( {\sqrt{2}, 0} \right)$ & $\sqrt{2}$ & $1$ & $\sqrt{2}$ 
&$m_{\text{s}}$  & 1 & $m_{\text{s}}$& $2$ 
\\ \hline
\rowcolor{pastelyellow} 
 $\frac{V_{f_2}u}{s} + \frac{V_{h_0}s}{u^3} $ &   $\left( \frac{2 \sqrt{2}}{7}, \frac{\sqrt{6}}{7} \right)$ &  $\sqrt{\frac{2}{7}}$ &  $ 2$ & $2\sqrt{\frac{2}{7}}$ 
 & $M_{\text{Pl,11}}$  & $\frac{7}{3}$& $m_{\text{KK,7}}$& $\frac{2}{3}$ 
 \\ \hline

$\frac{V_{f_4}}{s\,u} + \frac{V_{h_0}s}{u^3} $ & $\left( \frac{1}{\sqrt{2}}, \sqrt{\frac{3}{2}} \right)$ & $\sqrt{2}$ &  $\frac{1}{2}$ & $\frac{1}{\sqrt{2}}$ 
&$M_{\text{Pl,11}}$  & $\frac{5}{6}$& $m_{\text{KK,M-th}}$& $\frac{4}{3}$   
\\ \hline
 
$\frac{V_{f_6}}{s\,u^3} + \frac{V_{f_2}u}{s}$ & $\left( \sqrt{2}, 
  0 \right)$ & $\sqrt{2}$ & $1$ & $\sqrt{2}$ 
  & $m_{\text{s}}$  & 1 & $m_{\text{s}}$& $2$  
  \\ \hline

$\frac{V_{f_4}}{s\,u} $ & $\left( \sqrt{2}, \sqrt{\frac{2}{3}} \right)$ & $2\sqrt{\frac{2}{3}}$ & $\frac{3}{4}$ & $\sqrt{\frac{3}{2}}$ 
& $M_{\text{Pl,10}}$ & $\frac{3}{4}$&$m_{\text{KK,X}}$ &  $2$ 
\\ \hline

$\frac{V_{f_6}}{s\,u^3}$ & $\left(\sqrt{2}, \sqrt{6} \right)$ & $2\sqrt{2}$ & $\frac{1}{4}$ & $\frac{1}{\sqrt{2}}$ 
&$M_{\text{Pl,11}}$  & $\frac{5}{12}$& $m_{\text{KK,M-th}}$& $\frac{8}{3}$ 
\\ \hline
\end{tabular}
\caption{Different scalings along the possible asymptotic trajectories in the Type IIB  flux compactification at large complex structure and weak coupling, depending on which fluxes are turned on. The vector $\vec{\gamma}_{\text{gf}}$ is the projection of the $\zeta$-vector of the leading term of the potential \eqref{eq:potential1}, $V(s,u)\big|_{leading}$, along the gradient flow, while its length corresponds to the exponential decay rate of the potential along that trajectory. The coefficients of $V(s,u)\big|_{leading}$ are meant to be the ones given after axion stabilisation. $\rho_{\text{s}}$ is defined as $H \sim T_{\s}^{\rho_{\s}/2}$ as in \eqref{eq:rhos}, and $\alpha_{\text{s}}$ is the projection of the leading EFT string scale along the direction of the gradient flow as in \eqref{eq:alphas}. $\rho_{\text{sp}}$ is defined analogously for the species scale in \eqref{eq:rho-sp}. $\lambda$ is defined from $V \sim m_{\text{t}}^{\lambda}$ in \eqref{eq:lambdaV}. We highlight in yellow the case in tension with the TCC. The leading towers are given in terms of the mirror IIA microscopic description.}
\label{tab:IIBlargevollargecs}
\end{table}

\subsubsection*{Type IIB at weak string coupling away from large complex structure}
Let us now consider the F-theory regime corresponding to Sen-limit, but taking a different infinite distance limit in the Type IIB complex structure moduli space. In particular, we can consider the II$_{0,1} \to$ III$_{0,0}$ singularity enhancement in the CY$_4$ complex structure moduli space. The Kähler potential is 
\begin{equation}
    K(s,u) = - \log(s) - \log(u) \, ,
\end{equation}
such that the elementary EFT strings have tensions 
\begin{equation}
    T_{\text{s},1}(s,u) = \frac{1}{2s} \, ,  \quad T_{\text{s},2}(s,u) = \frac{1}{2u} \, ,
\end{equation}
in 4 dimensional Planck units and their respective $\zeta$-vectors are
\begin{equation}
    \zeta_{\text{s},1} = \left( \frac{1}{\sqrt{2}}, 0\right)\, ,  \quad \zeta_{\text{s},2} = \left( 0,\frac{1}{\sqrt{2}} \right)\, .  
\end{equation}

\begin{figure}[t]
    \centering
    \includegraphics[width=0.6\linewidth]{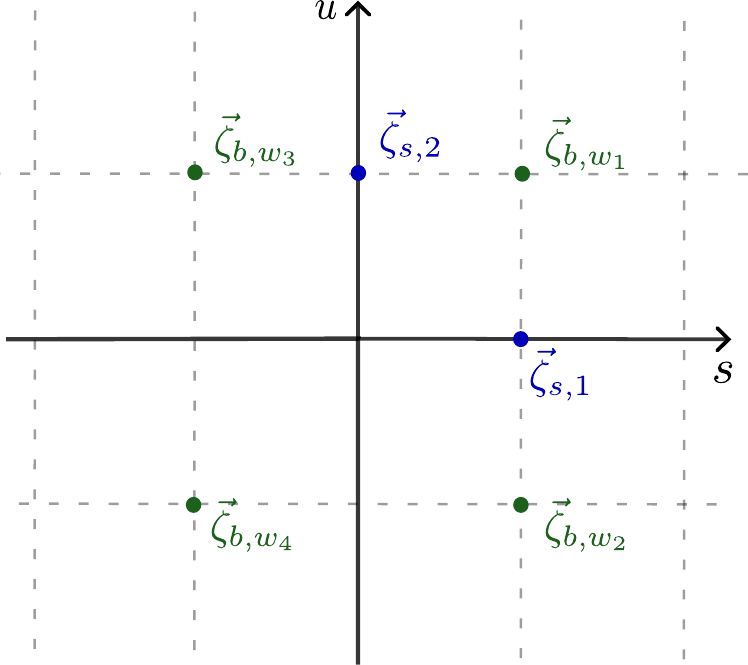}
    \caption{EFT strings (in blue) and EFT membranes (in green) for the EFT coming from Type IIB CY$_3$ orientifold at weak coupling.}
    \label{fig:bstypeIIB2}
\end{figure}

In this limit, the superpotential takes the form \cite{Grimm:2019ixq}
\begin{equation}
    W(\varphi^1,\varphi^2) = g_1 + g_2 \varphi^2 + \varphi^2 \left(g_3 + g_4 \varphi^2 \right) \, ,
\end{equation}
where the coefficients $g_i$ are related to the $G_4$ flux quanta. From \eqref{eq:branes}, the EFT membranes have tensions 
and $\zeta$-vectors given by 
\begin{align}
    T^0_{\text{b},g_1}(s,u) &= \frac{2 g_1}{\sqrt{su}} \, , &&\zeta_{\text{b},g_1} = \left( \frac{1}{\sqrt{2}},\frac{1}{\sqrt{2}} \right) \, , \\
    T^0_{\text{b},g_2}(s,u) &= 2 g_2 \sqrt{\frac{u}{s}} \, , &&\zeta_{\text{b},g_2} = \left( \frac{1}{\sqrt{2}}, - \frac{1}{\sqrt{2}} \right) \, , \\
    T^0_{\text{b},g_3}(s,u) &= 2 g_3 \sqrt{\frac{s}{u}} \, , &&\zeta_{\text{b},g_3} = \left(- \frac{1}{\sqrt{2}},\frac{1}{\sqrt{2}} \right) \, , \\
  T^0_{\text{b},g_4}(s,u) &= 2 g_4 \sqrt{su} \, , &&\zeta_{\text{b},g_4} = \left( -\frac{1}{\sqrt{2}},-\frac{1}{\sqrt{2}} \right) \, , 
\end{align}
and the $\zeta-$vector still live in the lattice of EFT strings, as can be seen in Figure \ref{fig:bstypeIIB2}. As in this asymptotic region the particle towers are not known, we show only the EFT string and membranes vectors.

The scalar potential is
\begin{equation} \label{eq:potential2}
    V(\varphi^1,\varphi^2) = \frac{V_{g_1}}{su} + \frac{V_{g_2}u}{s} + \frac{V_{g_3}s}{u} + V_{g_4} su+ V_{const} \, ,
\end{equation}
where again the coefficients $V_i$ depend on the axions and on the fluxes, and the subscript refers to the leading, axion-independent term in the same conventions as above.

We show in Table \ref{tab:IIB2} the gradient flow trajectories induced by this potential from \cite{Calderon-Infante:2022nxb} focusing on the fluxes that allow for vanishing asymptotic potential (i.e. $g_4=0$), together with the behaviour of the tension of the leading EFT string along any of these trajectories - though strictly speaking two of them are not admissible as the trajectory is not within the original growth sector. This limit does not show any clear pathology, as the potential does not source accelerated expansion, and at the same time the string is heavier than the would-be Hubble scale, as $\rho_{\s}<1$. 

\begin{table}[t]
    \centering
    \begin{tabular}{ |c|c|c|c|c|c| } 
 \hline
 Fluxes & $V(s,u)\big|_{\text{leading}}$ & $\vec{\gamma}_{\text{gf}}$ &$|\vec{\gamma}_{\text{gf}}|$ & $\rho_{\text{s}}$ & $\alpha_{\s}$ 
 \\  \hline
 $g_3= 0$ & $\frac{V_{g_2} u}{s}$ & $\left( \sqrt{2} , - \sqrt{2} \right)^*$ & $2$ & $\frac{1}{2}$ & 1  
 \\  \hline
 $g_2 =0$ & $\frac{V_{g_3} s}{u}$ & $\left(-\sqrt{2},\sqrt{2}\right)^*$ & $2$ & $\frac{1}{2}$ & 1 
 \\ \hline
 $g_2, g_3 = 0$ & $\frac{V_{g_1}}{s\, u} $ & $\left( \sqrt{2} , \sqrt{2}  \right)$ & $2$ & $\frac{1}{2}$ & 1 
 \\ \hline
\end{tabular}
    \caption{Different scalings along the possible asymptotic trajectories in the Type IIB  flux compactification away from large complex structure, depending on which fluxes are turned on. The vector $\vec{\gamma}_{\text{gf}}$ is the projection of the $\zeta$-vector of the leading term of the potential \eqref{eq:potential2}, $V(s,u)\big|_{leading}$, along the gradient flow, while its length corresponds to the exponential decay rate of the potential along that trajectory. The coefficients of $V(s,u)\big|_{leading}$ are meant to be the ones given after axion stabilisation. $\rho_{\text{s}}$ is defined as $H \sim T_{\s}^{\rho_{\s}/2}$ as in \eqref{eq:rhos}, and $\alpha_{\text{s}}$ is the projection of the leading EFT string scale along the direction of the gradient flow as in \eqref{eq:alphas}. The $*$ signals solutions that move us away from the growth sector.}
    \label{tab:IIB2}
\end{table}

\subsubsection*{F-theory away from Sen-limit}
\begin{figure}[t]
    \centering
    \includegraphics[width=0.6\linewidth]{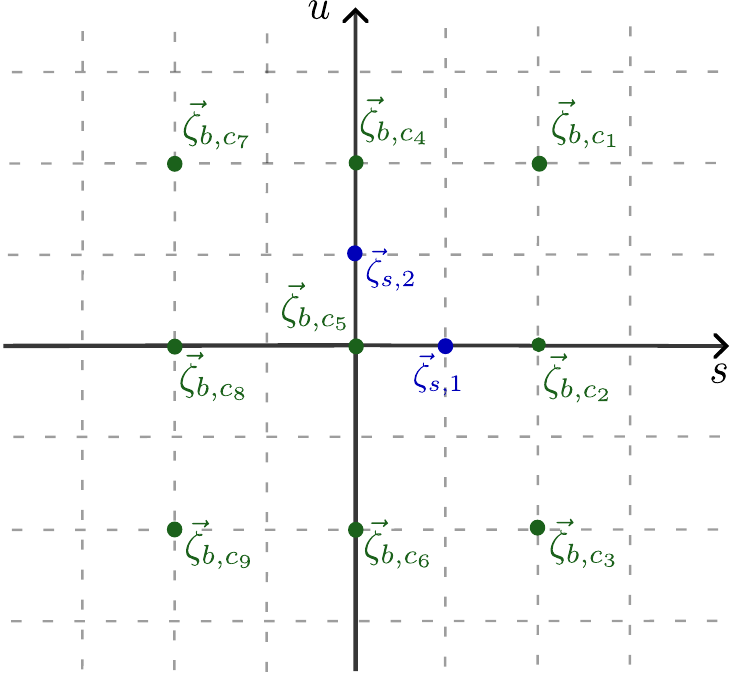}
    \caption{EFT strings (in blue) and EFT membranes (in green) for the EFT coming from the III$_{1,1} \to$ V$_{2,2}$ singularity enhancement of F-theory on a CY$_4$.}
    \label{fig:Fth}
\end{figure}

Let us now focus on the III$_{1,1} \to$ V$_{2,2}$ singularity enhancement in F-theory, which does not have a Type IIB construction as it is away from the weak string coupling regime. On this locus
\begin{equation}\label{eq:Fth-K}
    K(s,u) = - 2 \log(s) - 2 \log(u) \, ,
\end{equation}
such that the elementary EFT strings have tensions
\begin{equation}
    T_{\text{s},1} = \frac{1}{s} \, ,  \quad T_{\text{s},2} = \frac{1}{u} \, ,
\end{equation}
in 4 dimensional Planck units, and $\zeta$-vectors 
\begin{equation}\label{eq:EFTstrings-Fth}
    \zeta_{\text{s},1} = \left( \frac{1}{2}, 0\right)\, ,  \quad \zeta_{\text{s},2} = \left( 0,\frac{1}{2} \right)\, .  
\end{equation}
The superpotential can be expanded as \cite{Grimm:2019ixq}
\begin{align}
    W(\varphi^1,\varphi^2) &= c_1 + c_2 \varphi^2 + c_3 \left(\varphi^2\right)^2 +\varphi^1 \left( c_4 + c_5 \varphi^2 + c_6 \left( \varphi^2\right)^2 \right) \nonumber \\ &+ \left( \varphi^1 \right)^2 \left( c_7 + c_8 \varphi^2 + c_9 \left( \varphi^2 \right)^2\right) \, ,
\end{align}
and the EFT membranes in this limit have tensions and $\zeta$-vectors
\begin{align}
    T^0_{\text{b},c_1}(s,u) &= \frac{2 c_1}{su} \, , &&\zeta_{\text{b},c_1} = \left( 1,1 \right) \, , \\
    T^0_{\text{b},c_2}(s,u) &=  \frac{2 c_2}{s}  \, , &&\zeta_{\text{b},c_2} = \left( 1, 0 \right) \, , \\
    T^0_{\text{b},c_3}(s,u) &=   \frac{2 c_3 u}{s} \, , &&\zeta_{\text{b},c_3} = \left(1,-1 \right) \, , \\
    T^0_{\text{b},c_4}(s,u) &=   \frac{2 c_4}{u} \, , &&\zeta_{\text{b},c_4} = \left( 0,1 \right) \, , \\ 
    T^0_{\text{b},c_5}(s,u) &=  2c_5 \, , &&\zeta_{\text{b},c_5} = \left( 0,0 \right) \, , \\ 
    T^0_{\text{b},c_6}(s,u) &= 2c_6 u \, , &&\zeta_{\text{b},c_6} = \left( 0, - 1 \right) \, , \\
    T^0_{\text{b},c_7}(s,u) &=   \frac{2c_7 s}{u}  \, , &&\zeta_{\text{b},c_7} = \left(- 1,1 \right) \, , \\
    T^0_{\text{b},c_8}(s,u) &=   2c_8 s  \, , &&\zeta_{\text{b},c_8} = \left( -1,0 \right) \, , \\
    T^0_{\text{b},c_9}(s,u) &=  2c_9 su  \, , &&\zeta_{\text{b},c_9} = \left( -1,-1 \right) \, , 
\end{align}    
in 4 dimensional Planck units. We show the $\zeta$-vectors for EFT strings and membranes in Figure \ref{fig:Fth}. Again, the leading towers of particles are not known, and hence we do not display them here (see discussion below, though). 

The corresponding scalar potential reads
\begin{equation} \label{eq:Fthpotential}
    V(\varphi_1,\varphi_2)= \frac{V_{c_1}}{s^2u^2} + \frac{V_{c_2}}{s^2} + \frac{V_{c_3}u^2}{s^2} + \frac{V_{c_4}}{u^2} +  V_{c_6} u^2 + \frac{V_{c_7}s^2}{u^2} +  V_{c_8} s^2 + V_{c_9} s^2 u^2 + V_{const} \, ,
\end{equation}
and in order to select the terms that vanish asymptotically (at least along some trajectories) one must tune the fluxes in such a way that the coefficients $c_6, \, c_8$ and $c_9$ vanish. Moreover, notice that since this potential is symmetric under the exchange $u \leftrightarrow s$, we do not need to worry about the gradient flow trajectory going away from the growth sector we start with, as we had to do in Section \ref{sec:lightstrings} in order to consider the fully general situation.

The gradient flow trajectories and the relative scaling of the string tension with the Hubble scale along them are shown in Table \ref{tab:pureFtheory}, where we highlight the instances that are in tension with the strong dS conjecture (which are related by one another by the exchange $s \leftrightarrow u$). These are characterised by $|\vec{\gamma}_{\text{gf}}|=\frac{2}{\sqrt{5}}$, and an arrangement of the $\zeta$ vectors as in Figure \ref{fig:acc2} (which reflects the exchange symmetry of the potential, as the two convex hulls are equivalent under a rotation plus a reflection). We notice that these are precisely the cases where the string scale becomes of the same order as the Hubble scale.  Despite not having access to the nature of the string towers in these limits from the top-down, and hence not being able to spot any inconsistency in the hierarchy of particles and species scales, we can use the bottom-up reconstruction method of \cite{Grieco:2026yip} in order to check the allowed particles and species convex hulls for this moduli space, as we do in Appendix \ref{sec:app-Fth}. With this analysis reveal that the species scale and the tower would become lighter than the Hubble scale in the limit with acceleration expansion. 

Just as a final remark, notice that there is an example saturating the strong dS conjecture, $|\vec{\gamma}_{\text{gs}}|=\sqrt{2}$ that is completely fine in relation to the parameter $\rho_{\s}$, which is smaller than 1. There is no apparent inconsistency with the towers and species scale either, as considered in Appendix \ref{sec:app-Fth}.

\begin{table}[t]
    \centering
    \begin{tabular}{ |c|c|c|c|c|c|} 
 \hline
 Fluxes & $V(s,u)\big|_{\text{leading}}$ & $\vec{\gamma}_{\text{gf}}$ &$|\vec{\gamma}_{\text{gf}}|$ & $\rho_{\text{s}}$ &  $\alpha_{\s} $   
 \\  \hline
 All $c_i \ne 0$& $\frac{V_{c_1}}{s^ 2\, u^2}$ & $\left( 2,2 \right)$ & $2 \sqrt{2}$ & $ \frac{1}{4}$ & $\frac{1}{\sqrt{2}}$ 
 \\  \hline
 $c_5 \ne 0$& $\frac{V_{c_1}}{s^ 2\, u^2}$ & $\left( 2,2 \right)$ & $2 \sqrt{2}$ & $ \frac{1}{4}$ & $\frac{1}{\sqrt{2}}$  
 \\ \hline
\rowcolor{pastelyellow}$c_7,V_{const}=0$ & $\frac{V_{c_3}u^2}{s^2} +\frac{V_{c_4}}{u^2}   $ & $\left(  \frac{4}{5}, \frac{2}{5}\right)$ &  $\frac{2}{\sqrt{5}}$ &  $1$ & $\frac{2}{\sqrt{5}}$
\\ \hline
\rowcolor{pastelyellow}$c_3,V_{const}=0$ & $\frac{V_{c_2}}{s^ 2} + \frac{V_{c_7} s^2}{u^2}$ &  $\left( \frac{2}{5} , \frac{4}{5} \right)$ &  $\frac{2}{\sqrt{5}}$ &  $1$  & $\frac{2}{\sqrt{5}}$
\\ \hline
 $c_3,c_7=0$ & $\frac{V_{c_1}}{s^2 \, u^2} $ & $\left( 2,2 \right)$ & $2 \sqrt{2}$ &  $\frac{1}{4}$ & $\frac{1}{\sqrt{2}}$ 
 \\ \hline
$c_3,c_7,V_{const}=0$ & $\frac{V_{c_2}}{s^2} + \frac{V_{c_4}}{u^2} $ & $\left( 1,1 \right)$ & $\sqrt{2}$ & $\frac{1}{2}$ & $\frac{1}{\sqrt{2}}$ 
\\ \hline
 Only $c_1, c_2 \ne 0$ & $\frac{V_{c_2}}{s^2} $ & $\left( 2,0 \right)$ & $2$ & $\frac{1}{2}$ & 1 
 \\ \hline
Only $c_1, c_4 \ne 0$ & $\frac{V_{c_4}}{u^2}$ & $\left( 0,2 \right)$ & $2$ & $\frac{1}{2}$ & 1 
\\ \hline
Only $c_1 \ne 0$  & $\frac{V_{c_1}}{s^2\, u^2}$ & $\left( 2,2 \right)$ & $2\sqrt{2}$ & $\frac{1}{4}$ & $\frac{1}{\sqrt{2}}$ 
\\  \hline
    \end{tabular}
    \caption{Different scalings along the possible asymptotic trajectories in the F-theory  flux compactification away from Sen limit, depending on which fluxes are turned on. The vector $\vec{\gamma}_{\text{gf}}$ is the projection of the $\zeta$-vector of the leading term of the potential \eqref{eq:Fthpotential}, $V(s,u)\big|_{leading}$, along the gradient flow, while its length corresponds to the exponential decay rate of the potential along that trajectory. The coefficients of $V(s,u)\big|_{leading}$ are meant to be the ones given after axion stabilisation. $\rho_{\text{s}}$ is defined as $H \sim T_{\s}^{\rho_{\s}/2}$ as in \eqref{eq:rhos}, and $\alpha_{\text{s}}$ is the projection of the leading EFT string scale along the direction of the gradient flow as in \eqref{eq:alphas}.  In yellow, the solutions which violate the strong dS bound.}
    \label{tab:pureFtheory}
\end{table}

\subsubsection*{Type IIA on a CY$_3$ orientifold at weak string coupling and large volume} \label{sec:typeIIA}
In this section, we repeat the same analysis for the mirror-dual theory of the Type IIB CY$_3$ orientifold at large complex structure in Section \ref{sec:IIB1}, that is Type IIA orientifold of the mirror-dual Calabi-Yau threefold $\widetilde{\text{CY}}_3$ at large volume. The mirror duality leaves the 4d dilaton unchanged $\phi_{4}^{IIB} \leftrightarrow \phi_{4}^{IIA}$, while swapping the complex structure and the Kähler moduli. The Type IIA saxions $s$ and $t$ are given in terms of the 10 dimensional dilaton $\phi_{10}$, the overall volume in string units $\mathcal{V}_s$ and the complex structure $\Omega$  as follows (see e.g. \cite{Grimm:2019ixq,Calderon-Infante:2022nxb})
\begin{equation}
    s= e^{-\phi_{10}^{IIA}} \frac{\sqrt{\mathcal{V}_s^{IIA}}}{|\Omega^{IIA}|} \, , \quad t = (\mathcal{V}^{IIA}_s)^{\frac{1}{3}} \, .
\end{equation}
with $e^{\phi_4}=e^{\phi_{10}}/\mathcal{V}_s^{1/2}$. We denote the chiral complex structure modulus as $\varphi_1 = a_1 + i \, s$ and the Kähler one $\varphi^3 = a_3 + i \,t$.  The mirror map acts as 
\begin{equation}
    \phi_{4}^{IIB} \leftrightarrow \phi_{4}^{IIA}\, , \quad \mathcal{V}_s^{IIB} \leftrightarrow |\Omega^{IIA}|^2 \, , \quad |\Omega^{IIB}|^2 \leftrightarrow \mathcal{V}_s^{IIA} \, , 
\end{equation}
implying that the Kähler potential in IIA reads
\begin{equation} \label{eq:KIIA}
    K(s,u) = - 4\log \left(s\right) - 3 \log \left(t \right) \, .
\end{equation} 
From the Type IIA perspective, this corresponds to a slice moving along the 4d dilaton and volume directions, where the leading EFT strings are the fundamental string and an NS5 brane wrapping an even 4-cycle on the $\widetilde{\text{CY}}_3$, whose $\zeta-$vectors are respectively  \cite{Grieco:2025bjy}
\begin{equation}
    \zeta_{\text{s},1} = \left( \frac{1}{\sqrt{2}}, 0\right)\, ,  \quad \zeta_{\text{s},2} = \left( 0,\frac{1}{\sqrt{6}} \right)\, .  
\end{equation}

From mirror symmetry, the scalar potential can be obtained from \eqref{eq:potential1} as 
\begin{align} 
V^{\text{IIA}}(\varphi^1, \varphi^3) &= \frac{1}{s^3} V^{\text{IIB}}(\varphi^1, \varphi^2 \to \varphi^3) \nonumber\\ \label{eq:potentialIIA}
&=\frac{V_{f_6}}{s^4 t^3} + \frac{V_{f_4}}{s^4 t} + \frac{V_{f_2} t}{s^4} + \frac{V_{f_0} t^3}{s^4} + \frac{V_{h_0}}{s^2t^3} + \frac{V_{h_1}}{s^2 t} + \frac{V_{h_2}t}{s^2} + \frac{V_{h_3}t^3}{s^2} + \frac{V_{const}}{s^3} \, .
\end{align} 
The additional overall $\frac{1}{s^3}$ factor comes from the IIB overall factor that depends on the Einstein-frame volume, i.e. $\frac1{4(\mathcal{V}_E^{IIB})^2}=\frac{e^{3 \phi_{10}^{IIB}}}{4 (\mathcal{V}_s^{IIB})^2} \leftrightarrow \frac{1}{4s^3|\Omega^{IIA}|^4}$  \cite{Grimm:2019ixq,Calderon-Infante:2022nxb}. Analogously, the EFT membranes then have tensions and $\zeta$-vectors that can be obtained from \eqref{eq:tension1}-\eqref{eq:tension2} respectively as 
\begin{equation}
    T^{\text{IIA}}_{\b,i}(\varphi^1,\varphi^3) = \frac{1}{s^{3/2}}T^{\text{IIB}}_{\b,i}(\varphi^1,\varphi^2 \to \varphi^3) \, , \quad \zeta_{\text{b},i}^{\text{IIA}} = \left( \zeta_{\text{b},i}^{s,\text{IIB}}+k_{i},\zeta_{\text{b},i}^{u\to t,\text{IIB}}\right)\,,
\end{equation}
where $k_i=\frac{1}{\sqrt{2}}$ for R-R fluxes and $\sqrt{2}$ for NS-NS (including geometric \cite{Marchesano_2020}) fluxes.

In this setup, the towers are analogous to the ones in \eqref{eq:towers} by mirror symmetry. Moreover, in this IIA framework, we can provide a microscopic interpretation for them: 
\begin{itemize}
    \item $\vec{\zeta}_\text{p,1}=\left(\frac{1}{\sqrt{2}},0\right)$ is the scaling vector associated to the mass scale of a string oscillator tower $m_s$. Here, it is provided by the type II fundamental string;
    \item $\vec{\zeta}_\text{p,2}=\left(\frac{1}{\sqrt{2}},\frac{1}{\sqrt{6}}\right)$ is the scaling vector for a KK-6 tower decompactifying the full $\text{CY}_3$ manifold;
    \item$\vec{\zeta}_\text{p,3}=\left(0,\sqrt{\frac{3}{2}}\right)$ is the scaling vector of a KK-1 tower corresponding to the decompactification of a 1-cycle, which in this case is the M-theory interval.
\end{itemize}

The different gradient flow trajectories for the possible choices of fluxes as in \cite{Calderon-Infante:2022nxb} are shown in Table \ref{tab:IIAlargevol}, together with the rates of the leading EFT strings and particles along those directions. 
As can be seen in Table \ref{tab:IIAlargevol}, none of the physical trajectories yield accelerated expansion. Moreover, the tension of the string is always heavier than the scale of the scalar potential, and the leading particles have $\frac{18}{7} \leq \lambda \leq \frac{14}{3}$, which is consistent with the bounds in \eqref{eq:lambda1} and \eqref{eq:lambda2} -- which would combine to $\frac{1}{21} \leq \lambda \leq 14$. 

For the flux choices that are also considered in \cite{Casas:2024oak}, both the dS coefficients and the exponential rates of the leading towers agree with the results
there.\footnote{The regime considered in \cite{Casas:2024oak} (large
volume and weak coupling) contains the growth sector $\mathcal{R}_{st}$ as a subset -- such that it allows for the $h_0$-driven solution, that is not admissible within our growth sector.} The main discrepancy with their claim of $2 \leq \lambda \leq 4$ is that we allow for the flow driven by $f_6$ alone, with $\lambda=\frac{14}{3}$, that was not considered in \cite{Casas:2024oak} as it is argued to be always subleading with respect to the Casimir energy.

\begin{figure}
    \centering
    \includegraphics[width=0.4\linewidth]{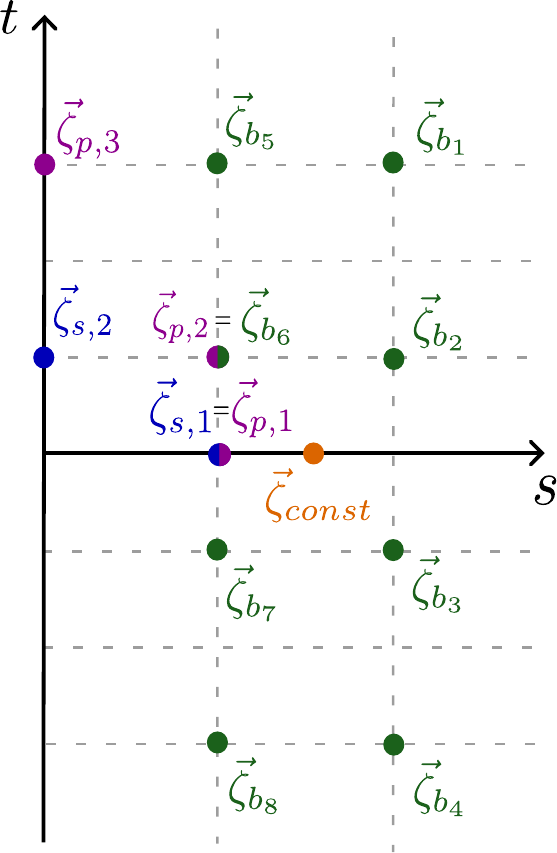}
    \caption{Particles (in purple), EFT strings (in blue) and EFT membranes (in green) for Type IIA CY$_3$ orientifold at weak coupling and large volume. In orange, the vector corresponding to $(V_\text{const})^{1/2}$ which is not associated to any brane and lives outside of the lattice.}
    \label{fig:IIApot}
\end{figure}

\begin{table}[ht!]
    \centering
    \begin{tabular}{ |c|c|c|c|c|c|c|c|c|} 
 \hline
 $V(s,t)\big|_{\text{leading}}$ & $\vec{\gamma}$ &$|\vec{\gamma}|$ & $\rho_{\s}$ & $\alpha_{\s}$ 
 & $\Lambda_{\text{sp}}$ & $\rho_{\text{sp}}$
 & Leading tower & $\lambda$\\  \hline
 $ \frac{V_{f_0}t^3}{s^4} + \frac{V_{h_0}}{s^2\,t^3}$ & $\left( \frac{18 \sqrt{2}}{13}, \frac{3\sqrt{6}}{13} \right)$ & $3 \sqrt{\frac{6}{13}}$ & $\frac{2}{3}$ &  $2 \sqrt{\frac{6}{13}}$ 
 & $m_{\text{s}}$& $\frac{2}{3}$ & $m_{\text{KK,X}}$ & $\frac{18}{7}$ \\  \hline
 $\frac{V_{const}}{s^3}$ & $\left( \frac{3}{\sqrt{2}},0 \right)$ & $\frac{3}{\sqrt{2}}$ & $\frac{2}{3}$ & $\sqrt{2}$ 
  & $m_{\text{s}}$& $\frac{2}{3}$ & $m_{\text{s}}$ & $3$ \\ \hline
 $\frac{V_{f_4}}{s^4 t} + \frac{V_{f_0}t^3}{s^4}$ & $\left( 2\sqrt{2},0 \right)$ & $2\sqrt{2}$ & $\frac{1}{2}$ & $\sqrt{2}$ 
  & $m_{\text{s}}$& $\frac{1}{2}$ & $m_{\text{s}}$& $4$ \\ \hline
 $\frac{V_{h_0}}{s^2 t^3} + \frac{V_{const}}{s^3}$ & $\left( \frac{18 \sqrt{2}}{13}, \frac{3\sqrt{6}}{13} \right)$ & $3\sqrt{\frac{6}{13}}$ & $  \frac{2}{3}$ &  $2 \sqrt{\frac{6}{13}}$ 
  & $m_{\text{s}}$ &$\frac{2}{3}$  & $m_{\text{KK,X}}$& $\frac{18}{7}$ \\ \hline
 $\frac{V_{f_2}t}{s^4} + \frac{V_{h_0}}{s^2\,t^3}$ & $\left( \frac{28\sqrt{2}}{19} , \frac{7 \sqrt{6}}{19} \right)$ & $7 \sqrt{\frac{2}{19}}$ &  $\frac{4}{7}$ & $4 \sqrt{\frac{2}{19}}$ 
  & $m_{\text{s}}$ &$\frac{4}{7}$  &$m_{\text{KK,X}}$ & $\frac{14}{5}$\\ \hline
 $\frac{V_{f_4}}{s^4 t} + \frac{V_{h_0}}{s^2t^3} $ & $\left( \frac{10 \sqrt{2}}{7} , \frac{5 \sqrt{6}}{7}\right)$ & $5 \sqrt{\frac{2}{7}}$ &  $\frac{2}{5}$ &  $2 \sqrt{\frac{2}{7}}$ 
 &$m_{\text{s}}$ & $\frac{7}{15}$ & $m_{\text{KK,7}}$& $\frac{10}{3}$\\ \hline
$\frac{V_{f_6}}{s^4 t^3} + \frac{V_{f_2}t}{s^4} $ & $\left( 2\sqrt{2},0 \right)$ & $2\sqrt{2}$ &  $\frac{1}{2}$ &  $\sqrt{2}$ 
 & $m_{\text{s}}$& $\frac{1}{2}$ &$m_{\text{s}}$ & $4$\\ \hline
$\frac{V_{f_4}}{s^4 t} $ & $\left( 2\sqrt{2}, \sqrt{\frac{2}{3}} \right)$ & $\sqrt{\frac{26}{3}}$ &  $\frac{6}{13}$ &  $2 \sqrt{\frac{6}{13}}$ 
 & $m_{\text{s}}$& $\frac{6}{13}$ & $m_{\text{KK,X}}$& $\frac{26}{7}$\\ \hline
 $\frac{V_{h_0}}{s^2 t^3} $ & $\left(  \sqrt{2}, \sqrt{6} \right)^*$ & $2\sqrt{2}$ &  $\frac{1}{4}$ &  $\frac{1}{\sqrt{2}}$ 
  & $M_{\text{Pl,11}}$&$\frac{5}{12}$  &$m_{\text{KK,M-th}}$ & $\frac{8}{3}$\\ \hline
 $\frac{V_{f_6}}{s^4 t^3} $ & $\left( 2\sqrt{2}, \sqrt{6} \right)$ & $\sqrt{14}$ &  $\frac{2}{7}$ &  $2 \sqrt{\frac{2}{7}}$ 
  & $M_{\text{Pl,11}}$& $\frac{1}{3}$  & $m_{\text{KK,7}}$ & $\frac{14}{3}$\\ \hline
\end{tabular}
    \caption{Different scalings along the possible asymptotic trajectories in the Type IIA  flux compactification at large volume and weak coupling, depending on which fluxes are turned on. The vector $\vec{\gamma}_{\text{gf}}$ is the projection of the $\zeta$-vector of the leading term of the potential \eqref{eq:potentialIIA}, $V(s,t)\big|_{leading}$, along the gradient flow, while its length corresponds to the exponential decay rate of the potential along that trajectory. The coefficients of $V(s,u)\big|_{leading}$ are meant to be the ones given after axion stabilisation. $\rho_{\text{s}}$ is defined as $H \sim T_{\s}^{\rho_{\s}/2}$ as in \eqref{eq:rhos}, and $\alpha_{\text{s}}$ is the projection of the leading EFT string scale along the direction of the gradient flow as in \eqref{eq:alphas}. $\rho_{\text{sp}}$ is defined analogously for the species scale in \eqref{eq:rho-sp}. $\lambda$ is defined as $V \sim m_{\text{t}}^{\lambda}$ as in \eqref{eq:lambdaV} . The $*$ signals solutions that move us away from the growth sector.}
\label{tab:IIAlargevol}
\end{table}

\subsection{Light strings from the top-down} \label{sec:4.1}

Despite the bottom-up analysis in Section \ref{sec:lightstrings} did not rule out accelerated expansion in the regimes of validity of the \n\, EFT, we have found that the specific scaling rates obtained in explicit string theory realisations forbid it; in all the examples violating either the strong dS conjecture and/or the TCC there are strings that become lighter than the Hubble scale. Let us elaborate on this.

First of all, we find that in III$_{1,1} \to$ V$_{2,2}$ singularity enhancement and in the \n\, Type IIA flux compactification there are no limits that are in tension with the strong dS conjecture or the TCC. At the same time, the corresponding EFT seem valid, as the strings remain heavy along all the admissible gradient flow trajectories. 

\begin{figure}
    \centering
    \includegraphics[width=0.6\linewidth]{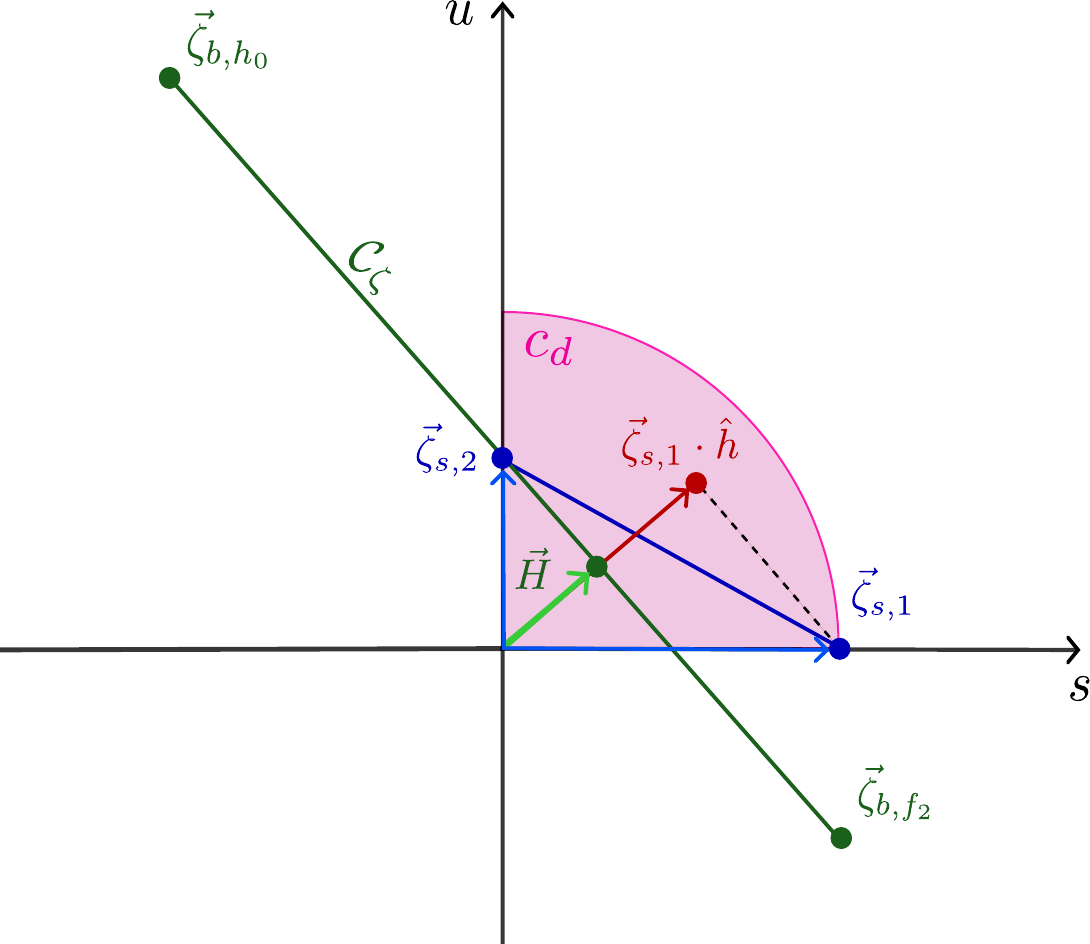}
    \caption{Branes and strings convex hulls respectively for the yellow row in Table \ref{tab:IIBlargevollargecs} compatible with accelerated expansion, i.e. with $|\vec{\gamma}_{\text{gf}}|=\frac{2}{\sqrt{7}}$. Along this limit, the Hubble scale (green arrrow)is heavier than the EFT string scale projected on the gradient flow (red arrow).}
    \label{fig:acc1}
\end{figure}

\begin{figure}
    \centering
    \includegraphics[width=0.5\linewidth]{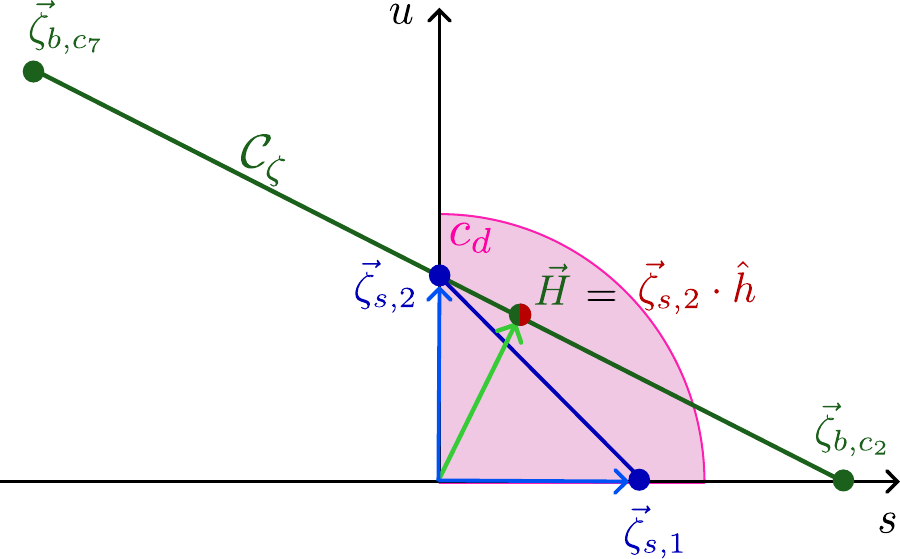}
    \includegraphics[width=0.3\linewidth]{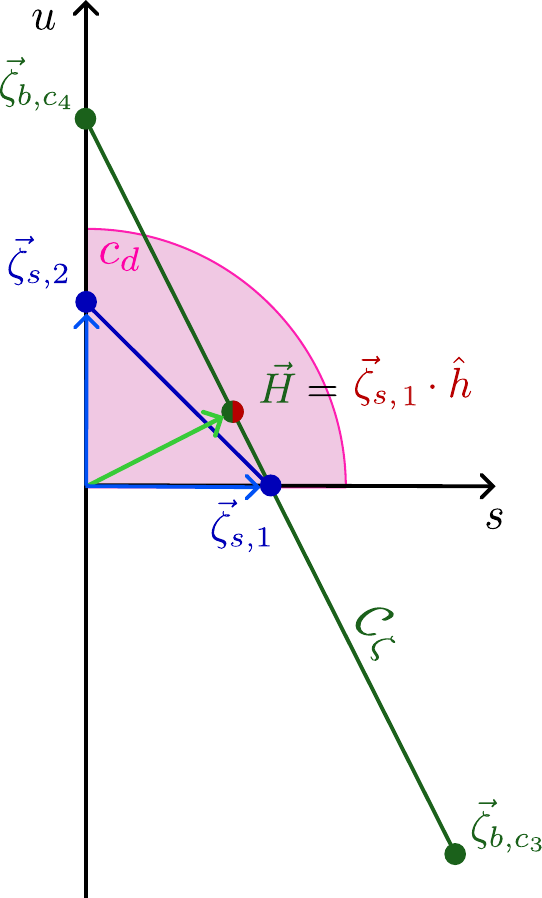}
    \caption{Branes and strings convex hulls respectively for the two instances in Table \ref{tab:pureFtheory} that are compatible with accelerated expansion, i.e. with $|\vec{\gamma}_{\text{gf}}|=\frac{2}{\sqrt{5}}$. Along this limit, the Hubble rate and the string tension vanish with the same rate.}
    \label{fig:acc2}
\end{figure}

Moving to the limits that are in tension with the aforementioned conjectures, on the other hand, we find that indeed there is always a string becoming too light, such that the theory we started with should receive corrections. In particular, the two examples apparently exhibiting accelerating expansion (found in \cite{Calderon-Infante:2022nxb}) are: 
\begin{itemize}
    \item II$_{0,1} \to$ V$_{2,2}$ singularity enhancement (large complex structure and weak coupling limit): despite being apparently well-controlled and at weak coupling, turning on at the same time non-trivial $f_2$ and $h_0$ fluxes results in accelerated expansion, since $\gamma = \sqrt{\frac{2}{7}}$ along that direction which is against the TCC, as was already noticed in \cite{Calderon-Infante:2022nxb} and shown here in Table \ref{tab:IIBlargevollargecs}. Notice that this example evades the no-go of \cite{Etheredge:2026cha}, since the NS flux term does not satisfy the conditions imposed for the brane scaling vectors in that paper. On the other hand, we see that the leading string gets tensionless twice as fast as the Hubble scale vanishes,  as is clear from the respective convex hulls intersecting as in  Figure \ref{fig:acc1}. This would invalidate the EFT description and signal that the supergravity action should receive corrections in this asymptotic regime (as also recently argued in \cite{Kaufmann:2026fli,Kaufmann:2026mha}). In Section \ref{sec:correctionsK} we argue that taking into account quantum corrections to divisor volumes, mixing the Kähler and complex structure limits in moduli space as proposed in \cite{Kaufmann:2026mha}, solves the issue of light EFT strings. At the same time, this forbids asymptotic accelerated expansion, with some caveats regarding the stabilization of the new quantum corrected volumes. 

    Analogously, in two out of three cases that would saturate the strong dS conjecture bound in table \ref{tab:IIBlargevollargecs}, we find that the leading EFT string becomes light parametrically with the same rate as the Hubble scale, so they are also susceptible to corrections. 
    \item III$_{1,1} \to$ V$_{2,2}$ singularity enhancement: as we mentioned around Table \ref{tab:pureFtheory}, in this limit there are two asymptotic trajectories that are incompatible with the strong dS conjecture -- both with $\alpha_\s = \frac{2}{\sqrt{5}}$. This also turns out to be a limit where the EFT description is no longer valid, as the string and the Hubble scale asymptote to zero at precisely the same rate (see Figure \ref{fig:acc2}), again suggesting the existence of corrections to the low energy supergravity action.  
\end{itemize}
In all the other gradient flow trajectories -- that are all passing the Swampland conjectures -- there are no light strings whatsoever, so there is no obvious obstruction to taking those limit from this perspective. Of course, this does not imply that these trajectories receive no corrections. If a particular flux choice requires a modification of the EFT in a given asymptotic regime, that modification will also affect other flux choices within the same regime. 


In the previous subsections, we also discussed the cases along which a tower of particle states appears to fall below the Hubble scale, posing a similar problem for the EFT description. Here, however, we have focused on EFT strings because their tensions can be determined directly from the low-energy EFT data, indicating that the low energy effective action requires modification. For particle towers, whose identification relies on additional UV input, we cannot exclude the possibility that further UV data resolve the apparent inconsistency.

Before concluding this subsection, let us remark that this explains what exactly goes wrong in the explicit asymptotic F-theory trajectories found in \cite{Calderon-Infante:2022nxb} apparently realizing accelerating expansion at parametric control. Those trajectories take place purely in the complex structure moduli space of F-theory flux compactifications, and ignore the K\"ahler moduli (assuming they can be stabilized by some other means). We can see now that the inconsistency of these trajectories can be directly inferred from the behaviour of the EFT strings in the complex structure moduli space, regardless of what happens with the K\"ahler moduli. This closes the top-down realization of the loophole for asymptotic acceleration found in \cite{Calderon-Infante:2022nxb}. It remains to be shown, though, that such bottom-up loophole (namely, having competing terms in the potential yielding a smaller slope) cannot be realized in some other way in string theory, and the underlying quantum gravity reason behind it.

\subsection{Quantum corrections, Kähler obstructions and accelerated expansion} \label{sec:correctionsK}

We have just shown that the candidate examples in which there would be asymptotically accelerated expansion are not under control at the effective theory level due to EFT strings becoming lighter than or of the same order as the Hubble scale. EFT strings are fundamental objects whose core cannot be resolved within the effective theory, so their mass scale should be above the UV cut-off and therefore above the Hubble scale, as explained around equation \eqref{eq:hierarchy}. This inconsistency can call for different resolutions. It could be that the asymptotic geometry of the complex structure moduli space and flux potential has to be corrected, such that either the infinite distance limit is obstructed or the mass hierarchy between the EFT string and Hubble scale is restored. In either case, this implies that there should be corrections to the asymptotic forms of the Kähler potential and flux superpotential describing the low energy EFT of F-theory on a $\text{CY}_4$  in section \ref{sec:topdown}. 

One might think that corrections of this kind should only become important as we move deeper into the interior of the complex structure moduli space for compactifications with \emph{simple} ingredients (only RR and NS fluxes). But these EFTs have very little supersymmetry, so the geometry of the field space does not share the same protection against quantum corrections as their analogous limits in 4d $\mathcal{N}=2$ CY compactifications. In a recent work \cite{Kaufmann:2026mha}, it has indeed been proposed that the Kähler potential should receive quantum corrections and that purely complex structure limits should be obstructed.

In this Section, we will consider a simple class of corrections to the Kähler potential motivated by those proposed in \cite{Kaufmann:2026mha} and check whether they are sufficient to either make the string heavier or to prevent accelerated expansion from happening, so that the hierarchy of scales \eqref{eq:hierarchy} would be restored.  

Consider F-theory compactified on an elliptically-fibered Calabi-Yau fourfold $W$ with fiber $\mathcal{E}$,
\begin{equation}\label{eq:fibration}
    \mathcal{E} \hookrightarrow W \to \mathcal{B}_3.
\end{equation}
In \cite{Kaufmann:2026mha}, the authors consider quantum corrections modifying the volume $\mathcal{V}_D$ of a divisor $D\in~\text{Eff}^1(\mathcal{B}_3)$ of the kind
\begin{equation} \label{eq:corrdivisor}
    \mathcal{V}^0_D \to \mathcal{V}_{D}^c = \mathcal{V}^0_{D} + c \, (\text{Im}\ z)^{\beta} \,, 
\end{equation}
when taking a suitable limit $\text{Im}(z)\to \infty$ in the complex structure moduli space.
The constant $c$ can be either positive or negative and depends on the specific details of the theory, such as the number of tensor multiplets. On the other hand, at least in the cases explicitly computed in \cite{Kaufmann:2026mha}, $\beta$ depends on the type of degeneration, more precisely $\beta=1,2$ for type II and III (infinite distance) singularities, respectively.
To maintain perturbative control on the $4$d \n action in the classical geometric setup while taking the large complex structure limit, it is necessary to co-scale the divisor classical volume  $\mathcal{V}^0_D\gtrsim  (\text{Im}\ z)^\beta$. This, in turn, implies that the pure complex structure infinite distance limit is obstructed in the quantum corrected moduli space. 

We are now interested in corrections to the Kähler potential for the type $\text{II}_{0,1}\to \text{V}_{2,2}$ (and $\text{III}_{1,1}\to \text{V}_{2,2}$) singularity enhancements. We have seen in section \ref{sec:topdown} that these classically correspond to type IIB at weak coupling and LCS, and a certain F-theory limit away from Sen-limit, respectively. These are the two setups where, in the classical moduli space, we find limits that admit accelerated expansion with light EFT strings that spoil the hierarchy between scales. Here we are considering two moduli  limits in $\text{CY}_4$ F-theory compactifications, for which computing general quantum corrections to the Kähler potential is a non-trivial task. Following the analysis of Kähler obstructions in section 4 of \cite{Kaufmann:2026mha}, we can put ourselves in a less general but more tractable setup where the leading corrections are known. 

\subsubsection*{Corrections to the $\text{II}_{0,1}\to \text{V}_{2,2}$ enhancement on a $\text{CY}_4$ moduli space}
Consider for simplicity a one-parameter Kähler moduli space with modulus $T=a_k+i\t$. In the 10-dimensional Einstein frame, the classical volume of $\mathcal{B}_3$ in \eqref{eq:fibration} is given by $\mathcal{V}_{\mathcal{B}_3}^0 \simeq t^{3/2}$, so that $t=\text{Im}\ T$ controls the volume of a divisor $D$. Recall that for a type IIB, weakly coupled and large complex structure regime the classical Kähler potential is given by

\begin{equation}
K_0= -2\log \mathcal{V}^0_{\mathcal{B}_3}+K_{cs},
\end{equation}
with $K_{cs}=-\log s -3 \log u$ as in \eqref{eq:IIBKahler}.

A quantum correction to the divisor $D$ of the form \eqref{eq:corrdivisor} is given by choosing $\beta=1$, motivated by the analysis of the standard Sen-limit in \cite{Kaufmann:2026mha}:
\begin{equation}
    \mathcal{V}_D^0 \simeq t \to \mathcal{V}_D^c \simeq t + cs = t_c.
\end{equation}
The corrected chiral coordinates are then
\begin{equation}\label{eq:coordchange}
    (t,s,u) \to (t_c,s,u),
\end{equation}
so, within our ansatz, the corrected Kähler potential is given by
\begin{equation}\label{eq:KcorrIIB}
    K_c = -3\log t_c  -\log s-3\log u.
\end{equation}
In this coordinate basis, it is straightforward to calculate the scalar potential $V$ for the moduli. Following \cite{Grimm:2019ixq, Calderon-Infante:2022nxb} and noticing that $K_c$ still respects a no-scale structure in the new chiral coordinates we can simply write, after axion stabilization,
\begin{equation}\label{eq:VcorrIIB}
\begin{split}
     V_c&= \frac{1}{(\mathcal{V}^c_{\mathcal{B}_3})^2}V_\text{cs}^\text{IIB}(s,u) = \\ &= \frac{1}{t_c^3}\left( \frac{V_{f_6}}{s u^3} + \frac{V_{f_4}}{s u} + \frac{V_{f_2} u}{s} +  \frac{V_{f_0} u^3}{s} + \frac{V_{h_0} s}{u^3} + \frac{V_{h_1}s}{u} + V_{h_2} su + V_{h_3} su^3 + V_{const}\right),
\end{split}
\end{equation}
where $V_\text{cs}^\text{IIB}(s,u) $ is given by \eqref{eq:potential1}. 

Elementary EFT strings in the corrected coordinates have tensions
\begin{equation}\label{eq:corrIIBtensions}
    T^c_t= -\frac{1}{2}\partial_{t_c}K_c=\frac{3}{2t_c}, \quad T^c_s=-\frac{1}{2}\partial_{s}K_c= \frac{1}{2s}, \quad T^c_u=-\frac{1}{2}\partial_{u}K_c= \frac{3}{2u}.
\end{equation}
Notice here a crucial point: $T_s^c$ has the same tension as the s-elementary EFT string $T_{s,1}$ in \eqref{eq:IIBtensions} (in the original coordinates) but it is not the same object. In particular, the EFT string flow is different because the EFT string charges are not the same for the two strings. In fact, EFT string charges $(e^0_t,e^0_s,e^0_u)$ transform under the coordinate change \eqref{eq:coordchange} as

    \begin{align}\label{eq:chargetr}
    \begin{pmatrix}
           e_t^c \\
           e_s^c \\
           e_u^c
         \end{pmatrix} =    \begin{pmatrix}
    1\quad & c\quad & 0 \\
    0\quad & 1\quad & 0 \\
    0\quad & 0\quad & 1
            \end{pmatrix}
\begin{pmatrix}
           e_t^0 \\
           e_s^0 \\
           e_u^0
         \end{pmatrix}.
  \end{align}
Hence, $T^c_{s}$ is not an elementary string in the old coordinates, having charges $(-c,1,0)$. This is compatible with the discussion in  section 4.1 of \cite{Kaufmann:2026mha} about the EFT string realizing the standard Sen-limit: if we want to preserve the F-theory supergravity description, it is necessary to take an additional infinite distance limit in the Kähler moduli space such that the EFT string sourcing the combined limit is a bound state of $T_{s,1}$ and the elementary EFT string along the classical volume direction, with tension $T_t^0 = \frac{3}{2t}$. This is exactly $T_s^c$. Moreover, it is compatible with the discussion about IIB/F-theory limits in \cite{Grieco:2025bjy}, where it was shown that the emergent type IIB string,  in the Kähler and axio-dilaton moduli space, does not coincide with the EFT string flow and is not BPS.

Regarding the original $T_{s,1}$ candidate string, its EFT string charges in the corrected coordinates are given by $(c,1,0)$ from \eqref{eq:chargetr}, so its tension is
\begin{equation} \label{eq:ORstring}
    T_{s,1}(t_c,s,u) = \frac{3c}{2t_c}+\frac{1}{2s}.
\end{equation}
Since $t_c=t+cs$, following this candidate EFT string flow yields different interpretations of the breakdown of the classical description depending on the sign of $c$. Recall that $T_{s,1}$ is the candidate EFT string whose flow corresponds to moving only toward $s\to \infty$ while keeping the classical volume $t=t_\star$ fixed. If $c<0$, the EFT string flow would be $t_{c}(\sigma) = t_\star-\abs{c}s$, which decreases and eventually leads to a domain in which the tension \eqref{eq:ORstring} blows up, so the flow breaks down. This implies that it is not truly an EFT BPS string, so the hierarchies of masses imposed above no longer apply. If $c>0$, the candidate flow is a priori under control, and we should check if there exists a gradient flow direction along which it becomes lighter than the Hubble scale. However, since it is a bound state of $T_{t}^c$ and $T_s^c$ in \eqref{eq:corrIIBtensions}, it will always be heavier than the elementary EFT strings and it is enough to check for these to be heavy enough to automatically guarantee that $T_{s,1}$ is.

Thanks to the no-scale structure, we have a complete factorization of the quantum corrected Kähler and complex structure sectors in \eqref{eq:VcorrIIB}. This means that each brane in  \eqref{eq:tension1} to \eqref{eq:tension2} simply acquires a dependence on $t_c$ of the kind
\begin{equation}
   T^\text{IIB,0}_{b}(s,u) \to  \frac{T^\text{IIB,0}_{b}}{t_c^{3/2}},
\end{equation}
and its scaling vector in this three-dimensional space is simply
\begin{equation}
    \zeta^\text{IIB}_c = \left(\sqrt{\frac{3}{2}}, \ \zeta^\text{IIB}_{0,1}, \ \zeta^\text{IIB}_{0,2}\right).
\end{equation}
Here, $\zeta^\text{IIB}_{0,1}$ and $\zeta^\text{IIB}_{0,2}$ are given by the first and second components of \eqref{eq:tension1} to \eqref{eq:tension2}.
It is straightforward to see that each gradient flow direction for a single term dominating in the potential will then have a Hubble scale given by
\begin{equation}\label{eq:correctedzmod}
    \abs{\zeta_\text{H}}=\abs{\zeta^\text{IIB}_c} = \sqrt{\frac{3}{2}+\abs{\zeta^\text{IIB}_{0}}^2},
\end{equation}
so that there will not be an asymptotically accelerating limit exactly as in the pure complex structure case with only one dominating term. However, this will now also be the case for two competing terms, with the Kähler modulus contribution in \eqref{eq:correctedzmod} being the same for every brane. In this case, in fact,
\begin{equation}
    \abs{\zeta_\text{H}}= \sqrt{\frac{\gamma^2}{4}+\frac{3}{2}}> \frac{1}{\sqrt{2}},
\end{equation}
where $\gamma$ is given by the entries in the third row of Table \ref{tab:IIBlargevollargecs}.
Moreover, from \eqref{eq:corrIIBtensions} we can see that the elementary EFT strings $\zeta$-vectors will have at most norm $\abs{\zeta_\text{str}}=\frac{1}{\sqrt{2}}$, hence the EFT strings will always be heavier than the Hubble scale. 
The co-scaling of the Kähler sector is hence responsible for the absence of acceleration and at the same time guarantees the EFT strings to be heavier than the Hubble scale. This is compatible with the discussion in \cite{Calderon-Infante:2022nxb} on how avoiding the contribution from the Kähler sector is crucial to have accelerated expansion, either by stabilizing it or by adding a suitable potential that would reduce its slope.

One could wonder whether there could be any way to stabilize the quantum corrected Kähler volume $t_c$ while still moving towards infinite distance in the complex structure/dilaton moduli space. In that case, we would recover the same setup we had before (upon exchanging $t\leftrightarrow t_c$),  allowing for accelerated expansion. However, this would be at the cost of having once again the problem of elementary EFT strings lighter than the Hubble scale. This is a direct consequence of the fact that the accelerated expansion is a product of moving only (or mostly) along the moduli that appear in the flux superpotential,  and does not really depend on the form of the corrections to the remaining moduli in the K\"ahler potential, as long as the latter do not play a relevant role in the gradient flow trajectory.
This suggests that either there is an obstruction to stabilizing $t_c$ while still moving on complex structure/dilaton directions, or the flux superpotential should also get corrected in a way that prevents asymptotic acceleration or resolves the hierarchy issue with the EFT strings.

Regarding the $\text{III}_{1,1}\to \text{V}_{2,2}$ singularity enhancement, we see that while the corrections for the volume divisors are different from the one considered for the Sen-limit in type IIB, the conclusions coincide. In particular, for a type III singularity we would expect a correction to a divisor volume $\mathcal{V}_D^0$ of the kind discussed in section 5 of \cite{Kaufmann:2026mha}, i.e. setting $\beta=2$ in \eqref{eq:corrdivisor}. Again, the absence of acceleration does not depend on the shape of the correction and co-scaling the Kähler sector is enough to avoid acceleration and light EFT strings at the same time regardless of the degeneration or geometry. 

As a final remark, note that in this section we considered for simplicity  corrections to the classical volume only in the $s$ direction, even if we are indeed considering a two-moduli $(s,u)$ complex structure space. In principle, one should also  consider corrections to the volumes in $u$ too, so that $u$ will also appear inside the definition of $t_c$. But this will not change the conclusions, as they are independent of the concrete definition of $t_c$ as explained above, as long as the K\"ahler potential still takes the form in \eqref{eq:KcorrIIB}. \\

\textbf{Acknowledgements}: We thank Christian Aoufia, Jeroen Monnee, Tom Rudelius, Ignacio Ruiz, Timo Weigand, and Max Wiesner for useful discussions. The authors thank
the Spanish Research Agency (AEI) for support through the grants IFT Centro de Excelencia
Severo Ochoa CEX2020-001007-S and CEX2025-001574-S, PID2021-123017NB-I00, PID2024-156043NB-I00, and PID2025-172338NB-I00,  funded
by MCIN/AEI/10.13039/ 501100011033 and by ERDF A way of making Europe. The research presented in this publication falls under the research line ``Strings and Quantum Gravity'' of CEX2025-001574-S. A.G. also thanks the Institute for Theoretical Physics in KU Leuven for hospitality during part of this work.
The work of I.V. (and V.C.) is supported by the ERC Starting Grant QGuide101042568 - StG 2021, and the Project ATR2023-145703 funded by MCIN/AEI/ 10.13039/501100011033. The work of A.G. is supported by the fellowship LCF/BQ/DI23/11990073 from ``La Caixa'' Foundation (ID 100010434), with additional funding from a STSM grant within the COST action CA22113 ``Fundamental Challenges in Theoretical Physics''.

\appendix
\section{F-theory away from Sen  limit: light towers and species scale} \label{sec:app-Fth}
In section \ref{sec:4.1} we have explored several top-down examples of 4d \n moduli spaces, given by type II/F-theory compactifications. In particular, we focused on the computation of scaling vectors for elementary EFT strings and branes, and only in the case of type IIA/B at weak coupling and large volume/large complex structure, scaling vectors for leading particles and species scale. To be precise, the light towers of states for the Kähler sector of type IIA Calabi-Yau orientifold compactifications are well known; see for example \cite{Lanza:2021udy, Grieco:2025bjy, Corvilain:2018lgw, Font:2019cxq, Castellano:2023jjt}. The particle content of type IIB $\text{CY}_3$ at large complex structure, as we discussed in \ref{sec:4.1}, is not known in its microscopic interpretation. However, we can make use of the mirror symmetry between its complex structure moduli space and the Kähler moduli space of type IIA on $\text{CY}_3$ to infer that the mass scales of light states have to be the same for both theories.
The mirror symmetry route is not available for type IIB away from large complex structure and from Sen's limit, and as of today, neither the mass scales of light towers nor their microscopic interpretation are well known for the resulting F-theory regime. 
There is a different route that can be taken towards understanding what the mass scales of F-theory light towers should be. In \cite{Grieco:2026yip}, it has been proposed that the SDC light towers content can be inferred, in 4d \n theories, just by the knowledge of elementary EFT strings and the compatibility of the candidate tower with the Emergent String Conjecture. The tension of an EFT string is completely fixed by the Kähler potential, which is just data specified by the 4d EFT. Hence, this procedure is bottom-up in nature and does not make reference to any top-down construction. Since the Kähler potential in the case of F-theory away from Sen-limit is just given by
\eqref{eq:Fth-K}, we can compute the corresponding EFT strings, draw the lattice they generate, and infer, following \cite{Grieco:2026yip}, the mass scale of candidate light towers for this strongly coupled regime. Given the exponential decay rates of the towers, we can infer what their interpretation as KK towers of string oscillator modes should be. It is, however, understood that this bottom-up procedure is not able to provide a microscopic interpretation of the mass scales in terms of stringy objects, nor  establish if they are populated by physical states.\footnote{In fact, taking into account that these limits might be obstructed by quantum corrections, as advocated in \cite{Kaufmann:2026mha} and in this paper, probably the towers do not exist. Here, we are only presenting what the towers should be from consistency with the swampland statements if such infinite distance limits existed. This will allow us to study the relative scaling between the towers and the potential, running into a contradiction that provides further evidence for the existence of quantum corrections obstructing the limits.}

Starting from these candidate scaling vectors  $\vec{\zeta}_{\text{p}}$ for light towers, we can then obtain the convex hull of species scale scaling vectors $\vec{Z}_{\text{sp}}$ by applying the pattern (for d=4) \cite{Castellano:2023jjt,Castellano:2023stg}
\begin{equation}\label{eq:pattern}  \vec{\zeta}_{\text{p}}\cdot \vec{Z}_{\text{sp}} = \frac{1}{2}.
\end{equation}

\begin{figure}
    \centering
    \includegraphics[width=0.48\linewidth]{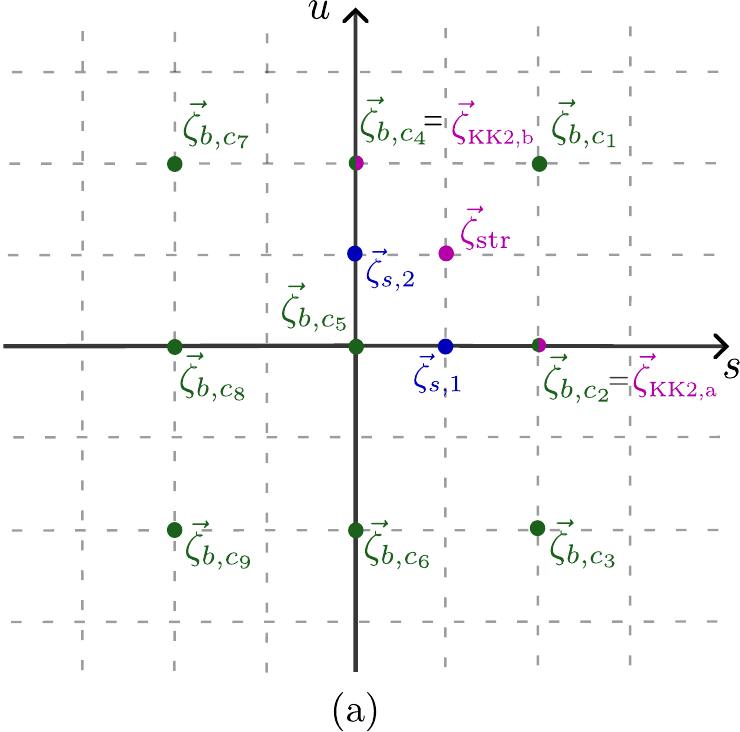}
    \quad
    \includegraphics[width=0.48\linewidth]{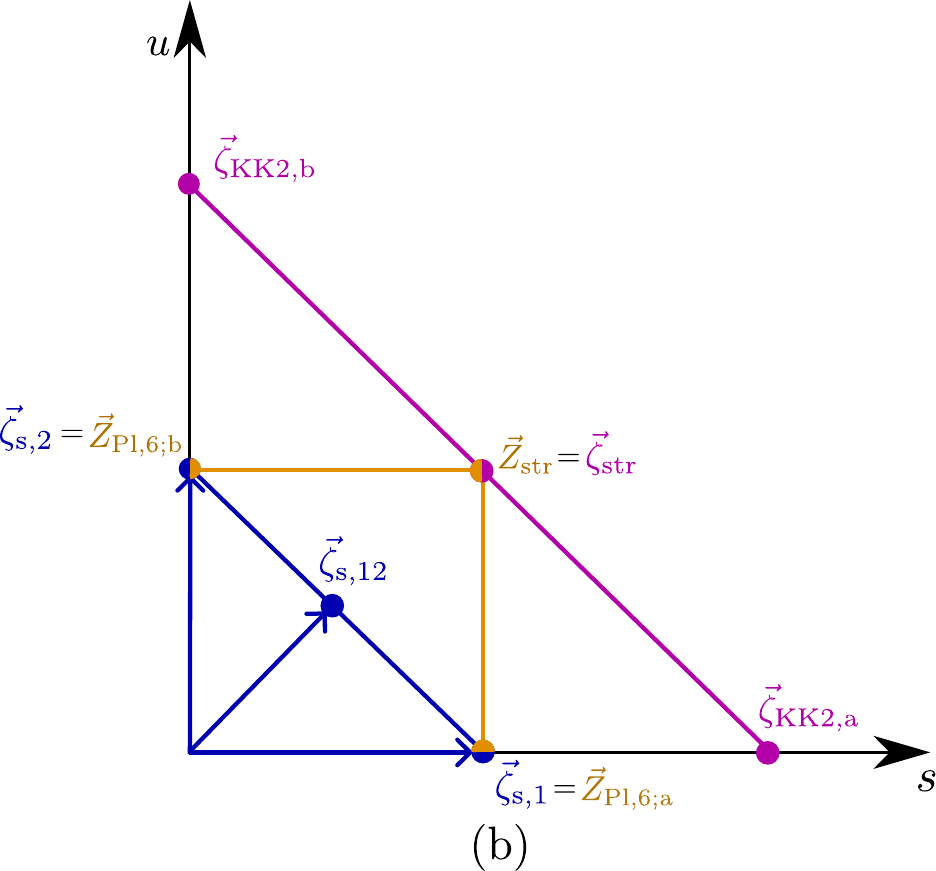}
    \caption{(a) EFT strings (in blue), EFT membranes (in green, and particles (in magenta) for the EFT coming from the III$_{1,1} \to$ V$_{2,2}$ singularity enhancement of F-theory on a CY$_4$. (b) Convex hulls of EFT strings (in blue), light towers of particles (in magenta), and species scale (in orange) for the EFT coming from the III$_{1,1} \to$ V$_{2,2}$ singularity enhancement of F-theory on a CY$_4$. Particles and species vector have been obtained through a bottom-up reconstruction following \cite{Grieco:2026yip}. \label{fig:Fth-hull-pot}}
   
\end{figure}

Following the above steps, we find that the only consistent convex hulls of the particle tower vectors and species vectors are those shown in Figure \ref{fig:Fth-hull-pot}. The elementary EFT strings scaling vectors in this setup are given by \eqref{eq:EFTstrings-Fth}.
Specifically, we have two leading towers for each saxionic direction that correspond to the decompactification of 2 dimensions and a bound state between the two with the exponential rate of an emergent non-BPS string. 
This is compatible with the tower structure that has been found in particular slices of the Kähler moduli space of M-theory compactified on a $G_2$ manifold \cite{Grieco:2025bjy}. There, the  perturbative regimes featuring such KK-2 towers could be interpreted as decompactification limits of the elliptical fiber of a $\text{CY}_4$, on which a dual F-theory is compactified. 

Hence, we have
\begin{equation}
    \vec{\zeta}_{\text{KK2,a}} = \left(1,0\right); \quad \vec{\zeta}_{\text{KK2,b}} = \left(0,1\right); \quad \vec{\zeta}_{\text{str}}=\left(\frac{1}{2},\frac{1}{2}\right).
\end{equation}

This translates through the pattern \eqref{eq:pattern} into the following species vectors:
\begin{equation}
    \vec{Z}_{\text{Pl,6;a}} = \left(\frac{1}{2},0\right); \quad \vec{Z}_{\text{Pl,6;b}} = \left(0,\frac{1}{2}\right); \quad \vec{Z}_{\text{str}}=\vec{\zeta}_{\text{str}}=\left(\frac{1}{2},\frac{1}{2}\right).
\end{equation}
In Table \ref{tab:pureFtheory-withparticles} we show, as an addition to Table \ref{tab:pureFtheory}, for each gradient flow trajectory the corresponding species scale (with its projection in units of $|\vec{\zeta}_\text{H}|$ given by $\rho_{\text{sp}}$) and leading towers with exponential rate $\lambda$. We can see that for the solutions in yellow, violating the strong dS bound, the leading EFT string and the Hubble scale are coincident ($\rho_\text{s}=1$), so there is no parametric scale separation between the two. The situation is actually worse than this if we look at the species scale: $\rho_\text{sp}=3/2$, so the hierarchy of scales required for an EFT to be under control is flipped. As observed in type II setups in section \ref{sec:topdown}, here too the Higuchi bound is not respected in the accelerating cases.

\begin{table}[t]
    \centering
    \begin{tabular}{ |c|c|c|c|c|c|c|c|c|} 
 \hline
  $V(s,u)\big|_{\text{leading}}$ & $\vec{\gamma}_{\text{gf}}$ & $|\vec{\gamma}_{\text{gf}}|$ & $\rho_{\text{s}}$ &  $\alpha_{\s} $ & $\Lambda_\text{sp}$ & $\rho_\text{sp}$ & Leading tower & $\lambda$ 
 \\  \hline
  $\frac{V_{c_1}}{s^ 2\, u^2}$ & $\left( 2,2 \right)$ & $2 \sqrt{2}$ & $ \frac{1}{4}$ & $\frac{1}{\sqrt{2}}$ & $m_\text{s}$& $\frac{1}{2}$ & $m_\text{s}$ & 4
 \\  \hline
\rowcolor{pastelyellow} $\frac{V_{c_3}u^2}{s^2} +\frac{V_{c_4}}{u^2}   $ & $\left(  \frac{4}{5}, \frac{2}{5}\right)$ & $\frac{2}{\sqrt{5}}$ &  $1$ & $\frac{2}{\sqrt{5}}$ &$m_\text{s}$& $\frac{3}{2}$ & $m_\text{KK2,a}$ & 1
\\ \hline
\rowcolor{pastelyellow}$\frac{V_{c_2}}{s^ 2} + \frac{V_{c_7} s^2}{u^2}$ &  $\left( \frac{2}{5} , \frac{4}{5} \right)$ &  $\frac{2}{\sqrt{5}}$ &  $1$  & $\frac{2}{\sqrt{5}}$ &$m_\text{s}$& $\frac{3}{2}$ & $m_\text{KK2,b}$ & 1
 \\ \hline
 $\frac{V_{c_2}}{s^2} + \frac{V_{c_4}}{u^2} $ & $\left( 1,1 \right)$ & $\sqrt{2}$ & $\frac{1}{2}$ & $\frac{1}{\sqrt{2}}$ & $m_{\text{s}}$& 1 & $m_\text{s}$ & 2
\\ \hline
 $\frac{V_{c_2}}{s^2} $ & $\left( 2,0 \right)$ & $2$ & $\frac{1}{2}$ & 1 & $M_{\text{Pl,6}}$ & $\frac{1}{2}$ & $m_{\text{KK2,a}}$ & 2
 \\ \hline
 $\frac{V_{c_4}}{u^2}$ & $\left( 0,2 \right)$ & $2$ & $\frac{1}{2}$ & 1 & $M_{\text{Pl,6}}$ & $\frac{1}{2}$ & $m_{\text{KK2,b}}$ & 2
\\ \hline
    \end{tabular}
    \caption{Different scalings along the possible asymptotic trajectories in the F-theory  flux compactification away from Sen limit, depending on which fluxes are turned on. The vector $\vec{\gamma}_{\text{gf}}$ is the projection of the $\zeta$-vector of the leading term of the potential \eqref{eq:Fthpotential}, $V(s,u)\big|_{leading}$, along the gradient flow, while its length corresponds to the exponential decay rate of the potential along that trajectory. The coefficients of $V(s,u)\big|_{leading}$ are meant to be the ones given after axion stabilisation. $\rho_{\text{s}}$ is defined as $H \sim T_{\s}^{\rho_{\s}/2}$ as in \eqref{eq:rhos}, and $\alpha_{\text{s}}$ is the projection of the leading string along the direction of the gradient flow. $\rho_{\text{sp}}$ is defined analogously for the species scale in \eqref{eq:rho-sp}. $\lambda$ is defined from $V \sim m_{\text{t}}^{\lambda}$ in \eqref{eq:lambdaV}. In yellow, the solutions which violate the strong dS bound.}
    \label{tab:pureFtheory-withparticles}
\end{table}

In conclusion, given the bottom-up reconstruction in terms of the EFT strings lattice for light towers and species scale, we can infer that not only there is no scale separation in the accelerating solutions for this strongly coupled F-theory limit, but actually the hierarchy of scales is inverted and we lose control of the EFT even more dramatically.

\section{Bounds on the potential-to-tower rate ratio} \label{app:B}

In this appendix we provide further details on the computation to obtain the bounds \eqref{eq:lambda1} and \eqref{eq:lambda2} on the $\lambda$ parameter, both in the presence of one and two leading terms in the potential.  

Let us start with just one term, where $\lambda$ is given as in \eqref{eq:lambdaVt} in terms of coefficients $\tilde{r}_i \in \mathbb{Z}_{\leq 0}$. The lower bound for such a $\lambda$ relies on the fact that $w_{i,\t}^{\text{max}}=3$, and that $|\tilde{r}_i| \geq 0$ and integer implies that 
\begin{equation}
    |\tilde{r}_i| (|\tilde{r}_i| - 1 ) \geq 0 \quad \implies \tilde{r}_i^2 \geq |\tilde{r}_i| \, .
\end{equation}
Hence
\begin{equation} \label{eq:lowerbound1}
    \lambda = \frac{2 \sum_i \tilde{r}_{i}^2/n_i}{ \sum |\tilde{r}_i|  w_{i,\t}/n_i} \geq \frac{2}{3} \frac{\sum_i |\tilde{r}_{i}|/n_i}{ \sum |\tilde{r}_i|/n_i} = \frac{2}{3} \, .
\end{equation}

The upper bound comes from the Integer Scaling Relation \eqref{eq:integral-scaling} together with the lattice condition $w_{i,\t}=0,1,2,3$. Together, these are equivalent to saying that the leading tower along $\vec{\zeta}_{s_i}$ -- call it $\vec{\zeta}_{\t^{(i)}}$ -- needs to have $w_{i,\t^{(i)}} \geq 1$.\footnote{If this did not happen, $\vec{\zeta}_{s_i}$ could not possibly be the leading tower in that direction, as $w_{i,\t} = 0$ and its mass would stay constant while moving in that direction.} Let us then consider this string, and compute
\begin{equation}
    \vec{\zeta}_{\t^{(i)}} \cdot \vec{\zeta}_V = \sum_j \frac{|\tilde{r}_j| w_{j,\t^{(i)}}}{n_j} \geq \frac{|\tilde{r}_i| w_{i,\t^{(i)}}}{n_i} \geq \frac{|\tilde{r}_i|}{n_i} \, ,
\end{equation}
where the first inequality holds because all the terms in the sum are non-negative. Then, if $\vec{\zeta}_{\t}$ is the leading tower along the gradient flow of the potential, it means that 
\begin{equation}
    \vec{\zeta}_{\t} \cdot \vec{\zeta}_V \geq \max_i \vec{\zeta}_{\t^{(i)}} \cdot \vec{\zeta}_V \geq \max_i \frac{|\tilde{r}_i|}{n_i} \, .
\end{equation}
Hence, since all the terms in the sum in the numerator are non-negative, we can write 
\begin{equation} \label{eq:upperbound1}
    \lambda = \frac{2 \sum \tilde{r}_i^2/n_i}{\vec{\zeta}_{\t} \cdot \vec{\zeta}_V} \leq \frac{2 \left(\max_j \frac{|\tilde{r}_j|}{n_j} \right) \sum |\tilde{r}_i|}{\max_i \frac{|\tilde{r}_i|}{n_i}} = 2 \sum_i |\tilde{r_i}| \leq 2 \sum_i n_i\, .
\end{equation}
The last inequality follows from footnote \ref{footnote4}. Hence, putting together \eqref{eq:lowerbound1} and \eqref{eq:upperbound1}
\begin{equation}
    \frac{2}{3} \leq \lambda \leq 2 \sum_i n_i \, .
\end{equation}

Let us now move to the case of two competing terms in the potential, expanded as  
\begin{equation} \label{eq:twotermsV}
    \vec{\zeta}_{V_1} = - 2 \sum_{i=1}^{N} \tilde{r}_{i,1} \vec{\zeta}_{\s_i} \, , \quad \vec{\zeta}_{V_2} = - 2 \sum_{i=1}^{N} \tilde{r}_{i,2} \vec{\zeta}_{\s_i} \, .
\end{equation}
These give rise to a gradient flow trajectory described by $\vec{\gamma}_{\text{gf}}$ that we can expand as
\begin{equation}
    \vec{\gamma}_{\text{gf}} \equiv - 2 \sum k_i \vec{\zeta}_{\s_i} \, ,
\end{equation}
where
\begin{equation} \label{eq:ki}
    k_i = \frac{(B-C)\tilde{r}_{i,1}+(A-C)\tilde{r}_{i,2}}{A+B-2C} \in \mathbb{Q} \, ,
\end{equation}
with $2A=|\vec{\zeta}_{V_1}|^2$, $2B=|\vec{\zeta}_{V_2}|^2$,  $2C=\vec{\zeta}_{V_1}\cdot\vec{\zeta}_{V_2}$. In order for the flow to be admissible, all the $k_i \leq 0$ with at least one non-vanishing. 

The lower bound is obtained as follows. First, since $w_{i,\t}^{\text{max}}=3$, it holds
\begin{equation}
    \lambda \geq \frac{|\vec{\zeta}_V|}{|\vec{\zeta_t}|} \geq \frac{\sqrt{2}}{3}\frac{|\vec{\zeta}_V|}{\sqrt{\sum_i \frac{1}{n_i}}}\, .
\end{equation}
Moreover
\begin{equation} \label{eq:lowerbound2}
    |\vec{\zeta}_V|^2 = \frac{2(A+B-C^2)}{A + B - 2C} \, , 
\end{equation}
and 
\begin{equation} \label{eq:denominator}
    A + B - 2C = \sum_i \frac{(\tilde{r}_{i,1}-\tilde{r}_{i,2})^2}{n_i} \leq 4 \sum_i \frac{\mathcal{R}_i^2}{n_i} \, ,
\end{equation}
as the biggest $(\tilde{r}_{i,1}-\tilde{r}_{i,2})$ is reached for $\tilde{r}_{i,2}=\tilde{r}_i^{\text{min}}$, and $\tilde{r}_{i,1} = \tilde{r}_i^{\text{max}}$, and the integer $\mathcal{R}_i$ is such that $-\mathcal{R}_i \leq \tilde{r}_i \leq \mathcal{R}_i$ -- from footnote \ref{footnote4}, $\mathcal{R}_i = \max(n_i+2d_{i-1},2d_i-n_i)$. Moreover
\begin{equation} \label{eq:numerator}
    A+B-C^2 = \sum_{i<j} \frac{(\tilde{r}_{i,1}\tilde{r}_{j,2}-\tilde{r}_{j,1}\tilde{r}_{i,2})^2}{n_i n_j} \geq \sum_{i<j} \frac{1}{n_i n_j} \geq \frac{1}{\max_{i<j}(n_in_j)} \, .
\end{equation}
In the first inequality, we used that at least one of the $\tilde{r}_{i,1}\tilde{r}_{j,2}-\tilde{r}_{j,1}\tilde{r}_{i,2} \ne 0$ being the two vectors $\vec{\zeta}_{V_1}$ and $\vec{\zeta}_{V_2}$ not parallel. By integrality, then at least one of these quantities needs to be $\geq 1$.

Putting \eqref{eq:lowerbound2}, \eqref{eq:numerator} and \eqref{eq:denominator} together
\begin{equation} \label{eq:mink}
    \lambda \geq \frac{1}{3 \sqrt{\max_{i<j}(n_in_j)\sum \frac{\mathcal{R}_i^2}{n_i} \sum \frac{1}{n_i}}}  \, .
\end{equation}
As for the upper bound, we can follow all the steps as in \eqref{eq:upperbound1} to get to $\lambda \leq 2 \sum_i |k_i|$, and moreover by the convex hull construction of the gradient flow vector there exists a $\hat{\tau} \in (0,1)$ such that 
\begin{equation}
    \sum_i |k_i| = (1-\hat{\tau}) \sum_i |\tilde{r}_{i,1}| + \hat{\tau} \sum_i |\tilde{r}_{i,2}| \leq \max(\sum_i|\tilde{r}_{i,1}|,\sum_i|\tilde{r}_{i,2}|) \leq \sum_i n_i \, ,
\end{equation}
so that the even the last inequality of \eqref{eq:upperbound1} holds for two potential terms. Together with \eqref{eq:mink}, this gives 
\begin{equation}
    \frac{1}{3 \sqrt{\max_{i<j}(n_in_j)\sum \frac{\mathcal{R}_i^2}{n_i} \sum \frac{1}{n_i}}} \leq \lambda \leq 2 \sum_i n_i \, , 
\end{equation} 
for two terms of the potential.

\bibliographystyle{JHEP}

\bibliography{biblio}

\end{document}